\documentclass[final,1p,11pt]{elsarticle}

\usepackage{epsfig}
\usepackage{array,tabularx,epsfig,mathrsfs,graphicx,rotating}
\usepackage{ifthen}
\usepackage{amsfonts}
\usepackage{ragged2e}
\PassOptionsToPackage{hyphens}{url}
\usepackage[hyphens]{url}
\usepackage{hyperref}
\usepackage{listings}
\usepackage{subfigure}
\usepackage{epstopdf}
\usepackage{color}
\usepackage{float}

\usepackage[normalem]{ulem} 
\usepackage{soul} 

\hypersetup{
  colorlinks=true,
  linkcolor=blue,
  citecolor=blue,
  urlcolor=blue
}

\graphicspath{{figs/}}

\newcommand{\beq}{\begin{equation}}
\newcommand{\eeq}{\end{equation}}

\chardef\til=126

\newcommand{\GEANTfour} {\textsc{geant4}}
\journal{ANL-HEP-132458, FERMILAB-PUB-17-064-CMS-E}

\begin{document}

\definecolor{mygreen}{rgb}{0,0.6,0} \definecolor{mygray}{rgb}{0.5,0.5,0.5} \definecolor{mymauve}{rgb}{0.58,0,0.82}

\lstset{ %
 backgroundcolor=\color{white},   
 basicstyle=\footnotesize,        
 breakatwhitespace=false,         
 breaklines=true,                 
 captionpos=b,                    
 commentstyle=\color{mygreen},    
 deletekeywords={...},            
 escapeinside={\%*}{*)},          
 extendedchars=true,              
 keepspaces=true,                 
 frame=tb,
 keywordstyle=\color{blue},       
 language=Python,                 
 otherkeywords={*,...},            
 rulecolor=\color{black},         
 showspaces=false,                
 showstringspaces=false,          
 showtabs=false,                  
 stepnumber=2,                    
 stringstyle=\color{mymauve},     
 tabsize=2,                        
 title=\lstname,                   
 numberstyle=\footnotesize,
 basicstyle=\small,
 basewidth={0.5em,0.5em}
}

\begin{frontmatter}

\title{
Initial performance studies of a general-purpose detector for multi-TeV physics at a 100~TeV $pp$ collider 
}

\author[add1]{S.V.~Chekanov}
\ead{chekanov@anl.gov}

\author[add1]{M.~Beydler}
\ead{mmbeydler@gmail.com}

\author[addDuke,add2]{A.V.~Kotwal}
\ead{ashutosh.kotwal@duke.edu}

\author[add2]{L.~Gray}
\ead{lagray@fnal.gov}

\author[addDuke]{S.~Sen}
\ead{sourav.sen@duke.edu}

\author[add2]{N.V.~Tran}
\ead{ntran@fnal.gov}

\author[add3]{S.-S.~Yu}
\ead{syu@cern.ch}

\author[addMSU]{J.~Zuzelski}
\ead{jwzuzelski18@gmail.com}

\address[add1]{
HEP Division, Argonne National Laboratory,
9700 S.~Cass Avenue,
Argonne, IL 60439, USA. 
}

\address[addDuke]{
Department of Physics, Duke University, USA
}

\address[add2]{
Fermi National Accelerator Laboratory
}

\address[addMSU]{
Department of Physics, Michigan State University, 220
Trowbridge Road, East Lansing, MI 48824 
}

\address[add3]{
Department of Physics, National Central University, Chung-Li, Taoyuan City 32001, Taiwan
}

\begin{abstract}
This paper describes simulations of detector response to multi-TeV particles and jets
at the Future Circular Collider (FCC-hh) or Super proton-proton Collider (SppC) which aim to collide proton beams with a centre-of-mass energy of 100 TeV.
The unprecedented energy regime of these future experiments
imposes new requirements on detector technologies which can be studied using
the detailed \GEANTfour\ simulations presented in this paper.
The initial performance of a detector designed for physics studies 
at  the FCC-hh or SppC experiments is described with an emphasis on
measurements of single particles up to 33~TeV in transverse momentum. 
The reconstruction of hadronic jets has also been studied in the transverse momentum range from 50~GeV to 26~TeV.
The granularity requirements for calorimetry  are investigated using the two-particle spatial resolution achieved for hadron showers.
\end{abstract}


\end{frontmatter}


\section{Introduction}

A 100~TeV proton-proton collider  leads to many challenges for detector design, and requires an optimized detector in order to achieve its physics goals. 
The capabilities of such a detector should include the ability to measure parameters of particles and jets
in the multi-TeV range. 
Such challenges exist for future circular $pp$ colliders of the European initiative, FCC-hh~\cite{Benedikt:2206376} and the Chinese initiative, SppC~\cite{Tang:2015qga}.

A promising starting point for this detector is provided by the Silicon Detector (SiD)
~\cite{Aihara:2009ad} concept, that was developed for the International Linear Collider (ILC)~\cite{Adolphsen:2013kya,Behnke:2013lya}.   
The SiD  is a compact,  general-purpose detector designed for high-precision  measurements of 
$e^+e^-$ collisions at a centre-of-mass energy of 500~GeV that can be 
extended to 1~TeV. 
The choice of silicon sensors for the tracking system and for the electromagnetic calorimeter ensures 
that the detector  
is resistant to beam backgrounds, while high-granularity calorimeters are well-suited for the  
reconstruction of individual particles and hadronic jets.
Key characteristics of the SiD detector are summarized in~\cite{Aihara:2009ad}.  
Together with efficient tracking, the fine segmentation of the calorimeter system optimizes the 
SiD detector for the use of particle-flow algorithms (PFA)  
which enables  identification and reconstruction of individual particles. 
The PFA objects can be reconstructed using the software algorithms implemented in 
the {\sc pandora} package~\cite{Charles:2009ta,Marshall:2013bda}.

The detector described and studied in this paper
is the Silicon Future Circular Collider (SiFCC) detector which shares many design features
with the SiD detector. This novel detector concept was studied using
detailed \GEANTfour\ simulation \cite{Allison2016186},  and is suitable for measuring 
particles and jets up to 30~TeV in transverse momentum ($p_{\rm T}$).

The unprecedented energy  of future experiments imposes new requirements on detector design.
The goal of software implementation of the SiFCC detector is to provide a versatile environment for simulations 
of detector performance, testing new technology options, event reconstruction techniques, as well as 
for assessment of the impact of specific detector designs on physics. 
Currently, several versions of the SiFCC detector 
are available for testing calorimeter and tracking technologies, providing an indispensable tool for the design of  the final version 
of the FCC-hh or SppC  detectors.

\section{Software implementation and Monte Carlo simulations}

The response of the SiFCC detector to physics processes
has been simulated using the Simulator for the Linear Collider (SLIC) software~\cite{Graf:2006ei}, which was 
developed for the ILC project.
The main strength of this software is the
easily configurable detector geometry using XML option files.  

One  approach for understanding detector effects  
on physics results is to use ``fast'' simulations of detector response, 
implemented in programs such as {\sc delphes}~\cite{deFavereau:2013fsa}, where detector resolution functions and 
efficiencies   are parameterized. In this strategy,  the impact of detector technology choices on physics is difficult to assess since the required response functions first need to be 
parameterized using fundamental principles. 
The approach used in this paper is based on the \GEANTfour\ toolkit with realistic detector description
based the  
Linear Collider Detector Description (LCDD). It also utilizes  
the SLIC software packages for realistic reconstruction of tracks and calorimeter clusters.   
These features  make these Monte Carlo simulations indispensable for
testing different detector designs from first principles.

 The truth-level samples used for detector simulation and
reconstruction 
were created using the ProMC package \cite{2013arXiv1311.1229C}.
The \GEANTfour\ simulation and event reconstruction based on SLIC software are described in~\cite{Chekanov:2016sme}.
The event samples are available from the
HepSim  database~\cite{Chekanov:2014fga}. 
All calculations, from event generation to simulation and reconstruction,     
were  performed using
the Open Science Grid~\cite{Pordes:2007zzb} and Argonne's Laboratory Computing Resource Center.
The creation of the event samples presented in this paper required one million CPU hours.

\section{Detector description}
\label{sec:detector}

\begin{figure}
\centering
\includegraphics[width=0.8\textwidth]{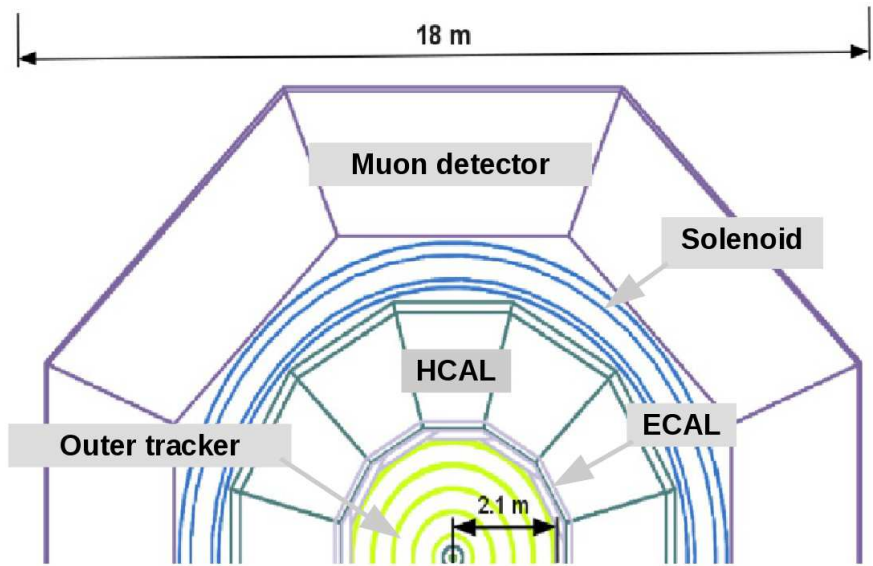}
\caption{The $x-y$ view of the SiFCC detector. The inner silicon tracker, with radius of 2.1~m, is surrounded by the electromagnetic and 
hadronic calorimeters inside the solenoid coil, and the muon detectors on the outside
are visible. The abbreviations shown in this figure are explained in Sect.~\ref{sec:detector}. 
}
\label{fig:sifcc11b}
\end{figure}

\begin{figure}
\centering
\includegraphics[width=0.8\textwidth]{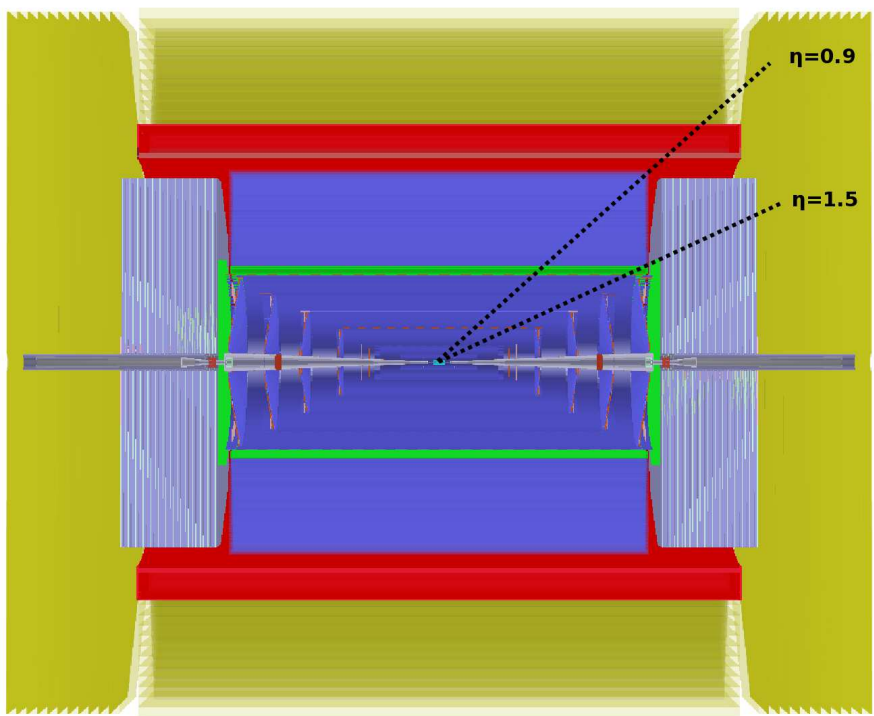}
\caption{The $r-z$ view of the SiFCC  detector. The solenoid that provides the 5~T magnetic field is shown in red. The 
electromagnetic calorimeter (ECAL) and the hadronic calorimeter (HCAL) are shown with the green and blue color, respectively. The light 
blue color from each side of the outer tracker is used to show the HCAL end-cap calorimeter. 
The inner silicon tracker inside the ECAL is shown in dark blue. The muon spectrometer
shown in yellow surrounds the solenoid magnet in the barrel region and the end-cap HCAL.
}
\label{fig:no_encap}
\end{figure}

The software implementation of the 
SiFCC detector leverages the original SiD detector design.
The size of the SiFCC detector has been  significantly extended.
The total length of the detector has been  increased to 20.1~m with an outer radius of 9~m.
Figure~\ref{fig:sifcc11a} illustrates the size comparison of the SiFCC detector with the original SiD detector, while
Fig.~\ref{fig:sifcc11b} shows the sizes of the SiFCC sub-detectors in the $x-y$ plane.
Figure~\ref{fig:no_encap} shows the $r-z$ view of the detector.

The main characteristics of the SiFCC detector are as follows, including specific differences between the SiFCC and SiD detectors:

\begin{figure}
\centering
\includegraphics[width=0.8\textwidth]{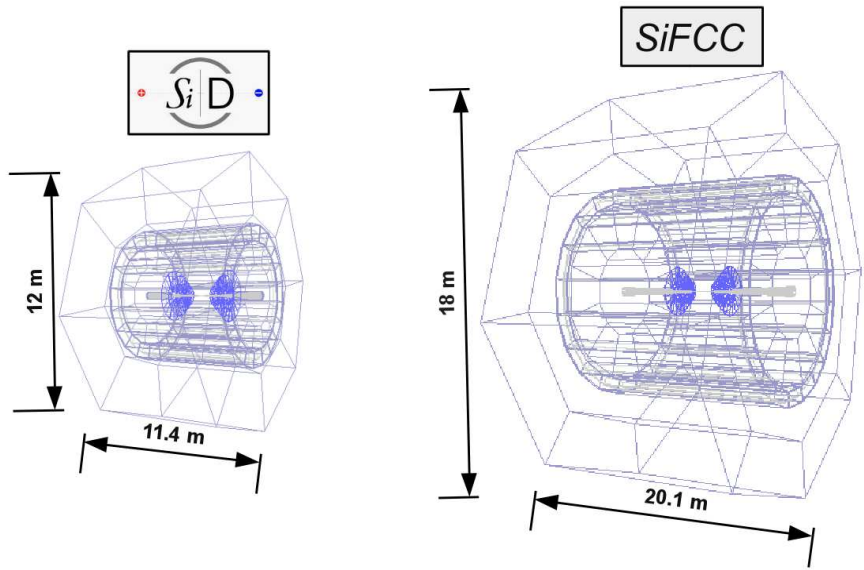}
\caption{A size comparison of the SiD and SiFCC detectors. See the text for the complete description and other differences.  
}
\label{fig:sifcc11a}
\end{figure}

\begin{figure}
\centering
\includegraphics[width=0.8\textwidth]{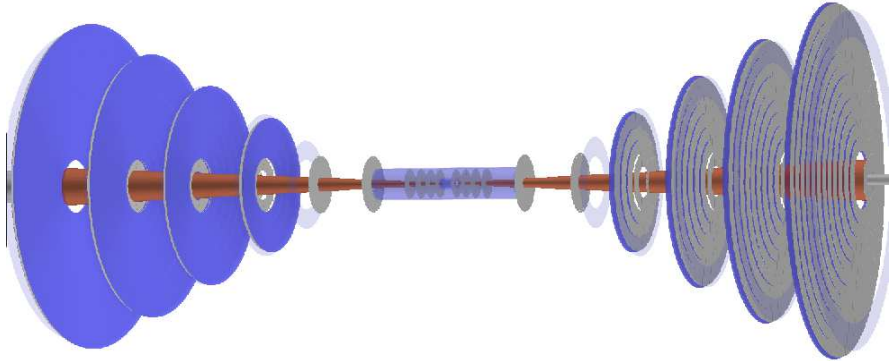}
\caption{Illustration of the pixel and microstrip forward disks of the silicon tracker of the SiFCC detector. 
The barrel microstrip modules are not shown. For improved  performance in the forward region,
the tracker was extended in the $z$-direction compared to the original SiD detector.}
\label{fig:det:stracking}
\end{figure}

\begin{itemize}

\item Almost $4\pi$ solid angle coverage for reconstructed particles.

\item A barrel tracker consisting of five layers of  silicon sensors with 50~$\mu$m  pitch.  
The forward tracker has four disks of silicon sensors.

\item Silicon pixel detector with 20~$\mu$m pitch,  consisting of five layers in the barrel and six disks in the
forward region.    
The pixel detector and the forward tracker are shown in Fig.~\ref{fig:det:stracking}.
The tracker was increased in size, and extended in the forward region, compared to the SiD detector in order to provide good measurements of tracks
in the $pp$ collision environment.
The tracking coverage up to a pseudorapidity\footnote{
As for many detectors designed for $pp$ collisions, the coordinate
 system of the SiFCC detector is a Cartesian right-handed coordinate system.
The nominal beam collision point is at the origin,  $(x,y,z)=(0,0,0)$.
The anti-clockwise beam direction around the collider defines the positive $z$-axis.
  The positive $x$-axis is defined as pointing radially outwards from the collider ring,  and the positive $y$-axis points upwards.
The azimuthal angle $\phi$ is measured around the beam
axis, and the polar angle $\theta$ is measured with respect to the $z$-axis.
The pseudorapidity
is given by $\eta=-\ln\tan(\theta/2)$. Transverse momentum is defined with respect  to the beam axis.}
of $|\eta|=3.5$ can be achieved.

\item Superconducting solenoid with a 5~T field extended in length from the original SiD
detector to provide a more uniform magnetic field in the forward region. Note that a 5~T field was also used in the  SiD design.

\item Highly segmented silicon-tungsten electromagnetic calorimeter (ECAL) with the transverse cell size of $2 \times 2$~cm. 
The ECAL has 30 layers built from tungsten pads with silicon readout, 
corresponding to 35~X$_{0}$. The first 20 layers use tungsten of 3~mm thickness. 
The electromagnetic sampling fraction is 1.47\%,  
as determined from single-photon and single-electron simulation samples by calculating the ECAL energy deposition in the active material.  The last ten layers use tungsten layers of  
twice the thickness, and thus have half the  sampling fraction.  There are two additional layers of sensors in front of the ECAL to 
serve as a pre-shower detector.  The calorimeter has $e/h$ close to $1.25$.

\item A steel-scintillator hadronic calorimeter (HCAL) with a transverse cell size of $5\times 5$~cm. The depth of the HCAL
in the barrel region is about 11.25 interaction lengths ($\lambda_I$).\footnote{Nuclear interaction length, $\lambda_I$,  is the average distance traveled by a hadron before undergoing an inelastic nuclear interaction.}  The increase in the total interaction lengths for an FCC 
 detector compared to the LHC detectors  was studied in~\cite{Carli:2016iuf}. The HCAL has 64 longitudinal layers in the barrel and the end-cap regions, which can be compared to the 40 layers of the original SiD detector. The sampling fraction of the HCAL  is $3.1\%$. 
Both ECAL and HCAL can reconstruct calorimeter clusters up to $|\eta|=3.5$. 

\item
The muon system located outside the solenoid which surrounds the HCAL calorimeter. The design of this detector   
closely follows the original SiD proposal~\cite{Aihara:2009ad} and  has an octagonal barrel geometry. The sensors cells are constructed using resistive plate chambers (RPC). 

\item
Silicon-tungsten beam calorimeter positioned 0.52~m away from the interaction point, along the beamline. 

\end{itemize}

For the studies presented below, track reconstruction  was performed with the LCSim package \cite{lcsim} using 
the seed tracker algorithm used for the SiD detector~\cite{Aihara:2009ad}. Seed tracker is a generic track-finding
algorithm based on a helix fitter. Tracks with $p_T > 500$~MeV were saved for final analysis. 

The \GEANTfour\ (version 10.2)~\cite{Allison2016186}  
was used for all results presented in this paper. As a check, low statistics event samples 
were created using \GEANTfour\ 10.3p1.  The
results were found to be consistent with the default \GEANTfour\ 10.2 used in this study.
The \GEANTfour\ simulation of the calorimeter response for inelastic processes is based
on the QGSP\_BERT physics list.  
The QGSP physics list,  based on the quark-gluon string model, is   
used for particles in the energy range 12~GeV -- 100~TeV.  
The validation of the QGSP physics list is available up to 400~GeV.
The FTFP physics list, based on the FRITIOF model,  is used for the energy range 9.5~GeV -- 25~GeV,
while the Bertini-Cascade model applies for particles below 9.9~GeV.
The elastic model ElasticHEP/Gheisha
is set to be valid up to 100~TeV. 
Discussions of these physics lists and models together with the references can be found in the \GEANTfour\ manual \cite{Allison2016186}.

For simulations of the calorimeter response, 
the sampling fraction of $3.1\%$  was used to correct hit energies in the HCAL,  and the sampling 
fractions of $1.47\%$ and $0.74\%$ were used 
in ECAL for the respective layers.
Selection cuts for calorimeter hits were applied based on the studies of the energy $E_{mip}$ of 
minimum ionizing particles (mip) for single muons with the energy of 100 GeV.
All hits with energy less than $E_{mip}-3\cdot RMS$, where $RMS$ is the root mean square of the mip distribution, were rejected.  
In addition, hits that have arrival time above 100~ns were rejected.  
No other readout-related affects (such as scintillator saturation at high energies per Birks' Law \cite{birks}) were included.

The calorimeter hits were clustered using a simple cone-based clustering algorithm~\cite{Thomson200925}. 
The PFA was not used for the performance studies presented in this paper since it requires optimization.

Tables~\ref{tab1:barrel} and~\ref{tab2:endcap} list the technology choices for each sub-detector of the SiFCC detector  
in the barrel and end-cap regions, respectively. 
Additional  parameters related to  various detector volumes can be found in  the HepSim repository~\cite{Chekanov:2014fga}.

\begin{table}[]
\centering
\caption{Technology and dimensions of the SiFCC sub-detectors in the barrel region. The solenoid field is given inside and outside the solenoid, respectively. }
\label{tab1:barrel}
\begin{tabular}{|l|l|r|r|r|}
Barrel          & Technology                & pitch/cell  &  radii  (cm) & $|z|$ size (cm) \\ \hline 
Vertex detector & silicon pixels/5 layers   & 25 $\mu$m  & 1.3 -   6.3      &   38        \\
Outer tracker   & silicon strips/5 layers   & 50 $\mu$m & 39   -  209         &   921       \\
ECAL            & silicon pixels+W     & 2$\times$2 cm  & 210       - 230    &  976  \\           
HCAL            & scintillator+steel   & 5$\times$5 cm & 230          - 470    &  980  \\
Solenoid        & 5~T (inner), -0.6~T (outer)    &  - & 480          - 560     &  976   \\ 
Muon detector   & RPC+steel   & 3$\times$3 cm & 570          -  903     &  1400  \\ 
\end{tabular}
\end{table}


\begin{table}[]
\centering
\caption{Technology and dimensions of the SiFCC sub-detectors for the end-cap region.}
\label{tab2:endcap}
\begin{tabular}{|l|l|r|r|r|}
End-cap          & Technology               & pitch/cell &  $z$ extent  & outer radius  \\ \hline
                &                          &                  &      (cm)        & (cm)      \\ \hline
Vertex detector & silicon pixels           &  25 $\mu$m  &                      &         \\
Outer tracker   & silicon strips           &  50 $\mu$m  &                      &          \\
ECAL            & silicon pixels+W         &   2$\times$2 cm    &   	500   - 516      & 250      \\           
HCAL            & scintillator+steel       &   5$\times$5 cm    &     	518   -	742       &  450       \\ 
Muon detector   & RPC+steel       &   3$\times$3 cm     &     745   - 1010      &  895      \\ 
Lumi calorimeter  &  silicon+W              &  3.5$\times$3.5~mm  &   	495   - 513       &  20      \\ 
Beam calorimeter  & semiconductor+W        &   3.5$\times$3.5~mm  &    520   - 539       & 13      \\ 
\end{tabular}
\end{table}

\begin{figure}
\centering
  \subfigure[$\eta$ distribution of tracks] {
  \includegraphics[scale=0.45]{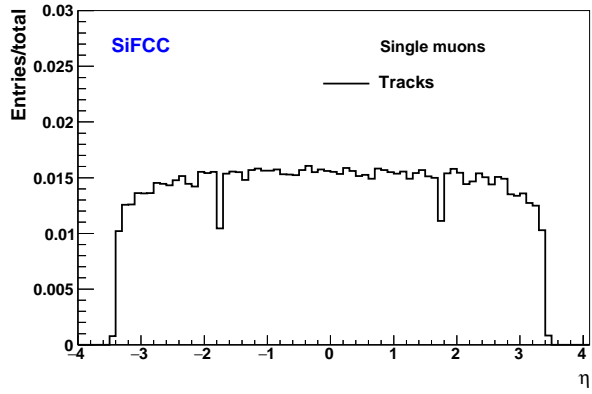}
  }
  \subfigure[$\eta$ distribution of reconstructed clusters] {
  \includegraphics[scale=0.45]{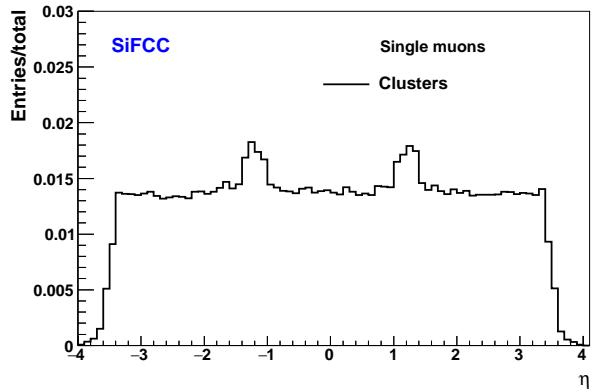}
  }
\caption{The pseudorapidity distributions of reconstructed tracks and calorimeter clusters in the SiFCC detector. Single muons with 
 uniform distributions in $\eta$ ($|\eta| < 4$) and $\phi$, and uniform distribution in momentum ($p$) in the range $4<p<1024$~GeV,  were used.}
\label{fig:eta}
\end{figure}

Figure~\ref{fig:eta} shows the pseudorapidity distributions of reconstructed tracks and calorimeter clusters. 
 The distributions were obtained for single muons which were generated with a uniform momentum distribution between 4~GeV and 1024~GeV, and
 uniform distributions in the $-4<\eta<4$ range  
 and in azimuthal angle. This figure shows the pseudorapidity range which can be used for
physics analyses. The enhancement near $|\eta|=1.2$ for calorimeters clusters is  due to the transition region between 
barrel and end-cap calorimeters, where the same object is reconstructed in both calorimeters and the clusters are not merged. 
The suppression  near $|\eta|=1.7-1.8$ for tracks is due to the transition from the barrel to the end-cap sub-detectors  of the silicon tracker.

\newpage
\section{Tracking performance}
\label{sec:tracking}

The track reconstruction efficiency as a function of track momentum ($p$) was investigated in the
pseudorapidity range $|\eta|<4$.
The study was performed  using single muons generated at fixed values of  $p = 2^n$~GeV, where $n=1, 2 ... 15$, after full detector simulation and reconstruction. 
The efficiency was calculated using reconstructed tracks matched to truth tracks with  
$\eta$-$\phi$ distance less than 0.15.  The following requirements were made on the reconstructed
 tracks to define efficiency: (1) The presence of at least six hits in the silicon pixel and microstrip layers; 
(2) The maximum distance of closest approach (DCA) $|{\rm DCA}|<6$~mm; (3) The longitudinal impact parameter, $z_0$, defined as the
difference between the $z$ coordinates of the primary vertex position to the track at this point of closest approach in $r-\phi$, is $|z_0|<10$~mm; (4)
$\chi^2 < 10$ for the track fit in the track reconstruction.     

\begin{figure}
\centering
\includegraphics[width=0.47\textwidth]{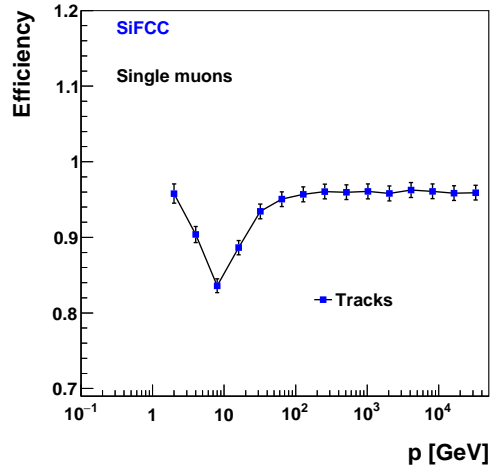}
\caption{Tracking efficiency as a function of track momentum, $p$, in the pseudorapidity range $|\eta|<4$.}
\label{fig:pap_effic_pt}
\end{figure}

Figure~\ref{fig:pap_effic_pt} shows the track reconstruction efficiency as a function  
of muon momenta, $p$. A dip in the efficiency near 8~GeV is related to
a lower efficiency of the forward (large $\eta$) regions for low-momentum tracks.

\begin{figure}
\centering
  \subfigure[$p=2$ GeV] {
  \includegraphics[width=0.46\textwidth]{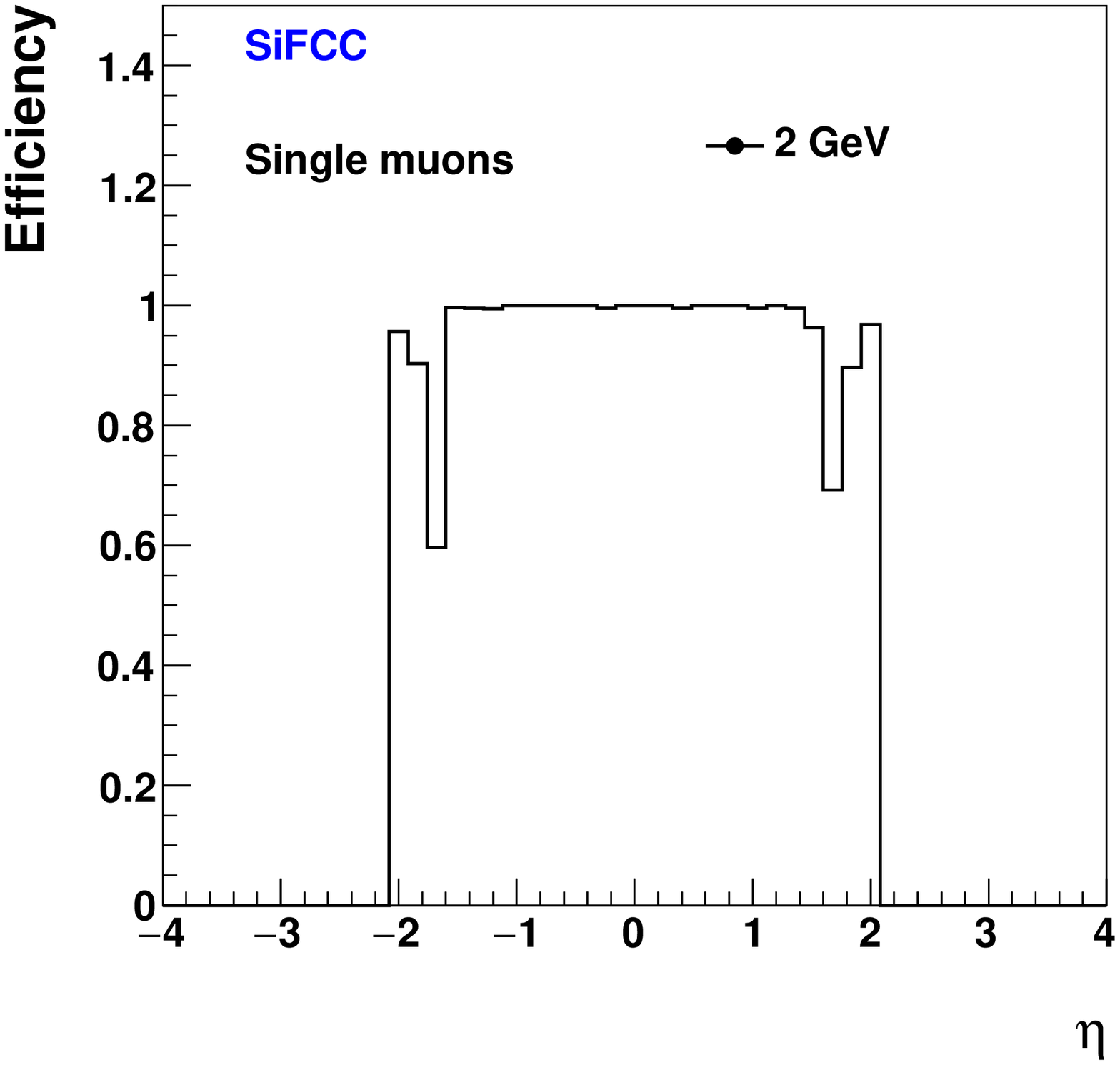}
  }
  \subfigure[$p=16.4$ TeV] {
  \includegraphics[width=0.46\textwidth]{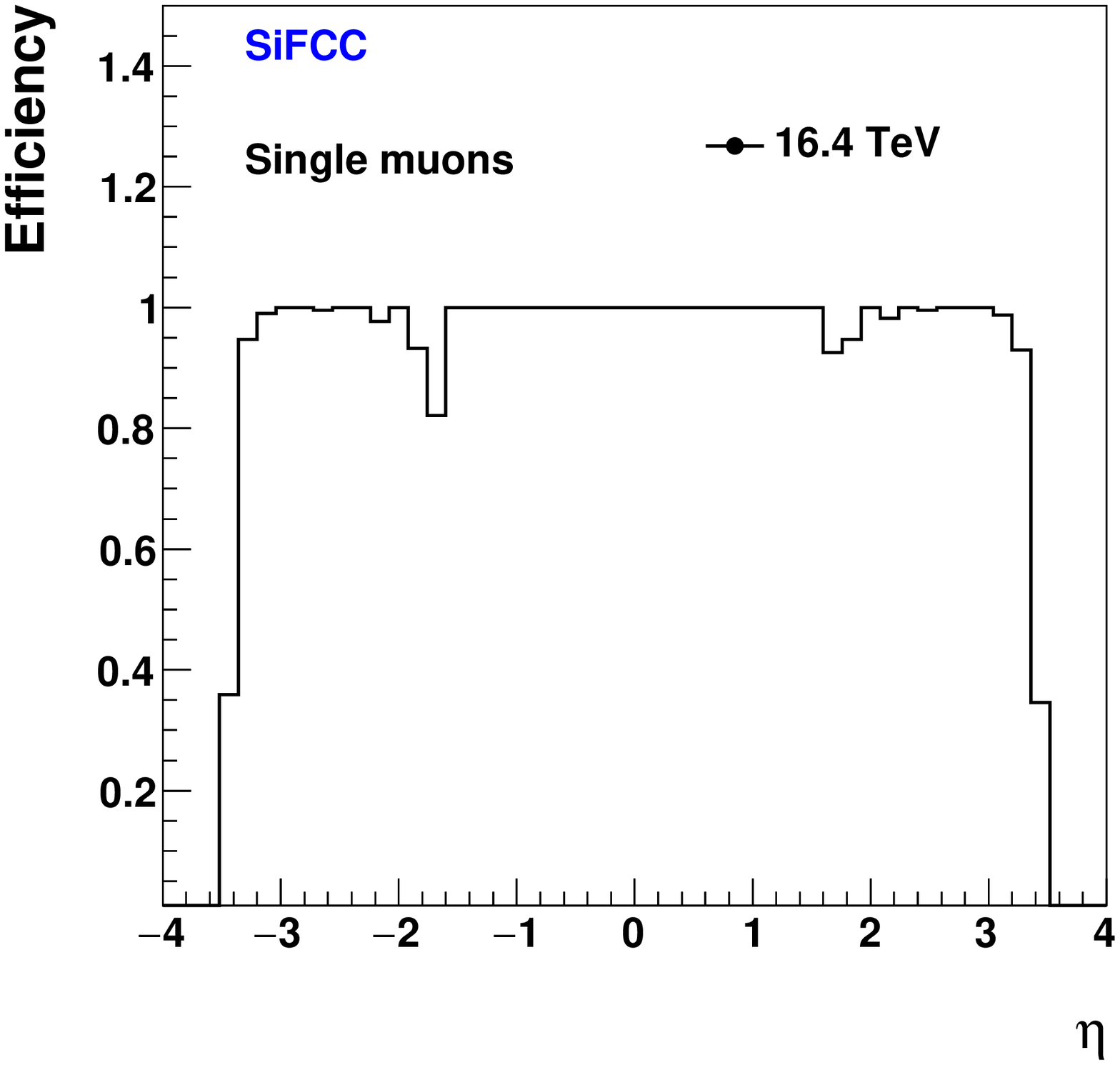}
  }
\caption{Tracking efficiency as a function of $\eta$ for two different values of muon momenta $p$.}
\label{fig:eff_eta}
\end{figure}

Figure~\ref{fig:eff_eta} shows the tracking efficiency as a function of $\eta$ for two different $p$ values. 
The efficiency is above $99\%$ in the central $\eta$ region.
However, tracks with momenta close to 2~GeV cannot be reconstructed in the forward region ($|\eta|>2$) 
because they are strongly curved by the  5~T magnetic field.
The drop in the efficiency 
in the region $|\eta|=1.6-1.8$ is due to the transition from the barrel to the end-cap sub-detectors  of the silicon tracker.

\begin{figure}
\centering
  \subfigure[$\sigma(p_T)/p_T$ vs $p$ ]{
  \includegraphics[width=0.46\textwidth]{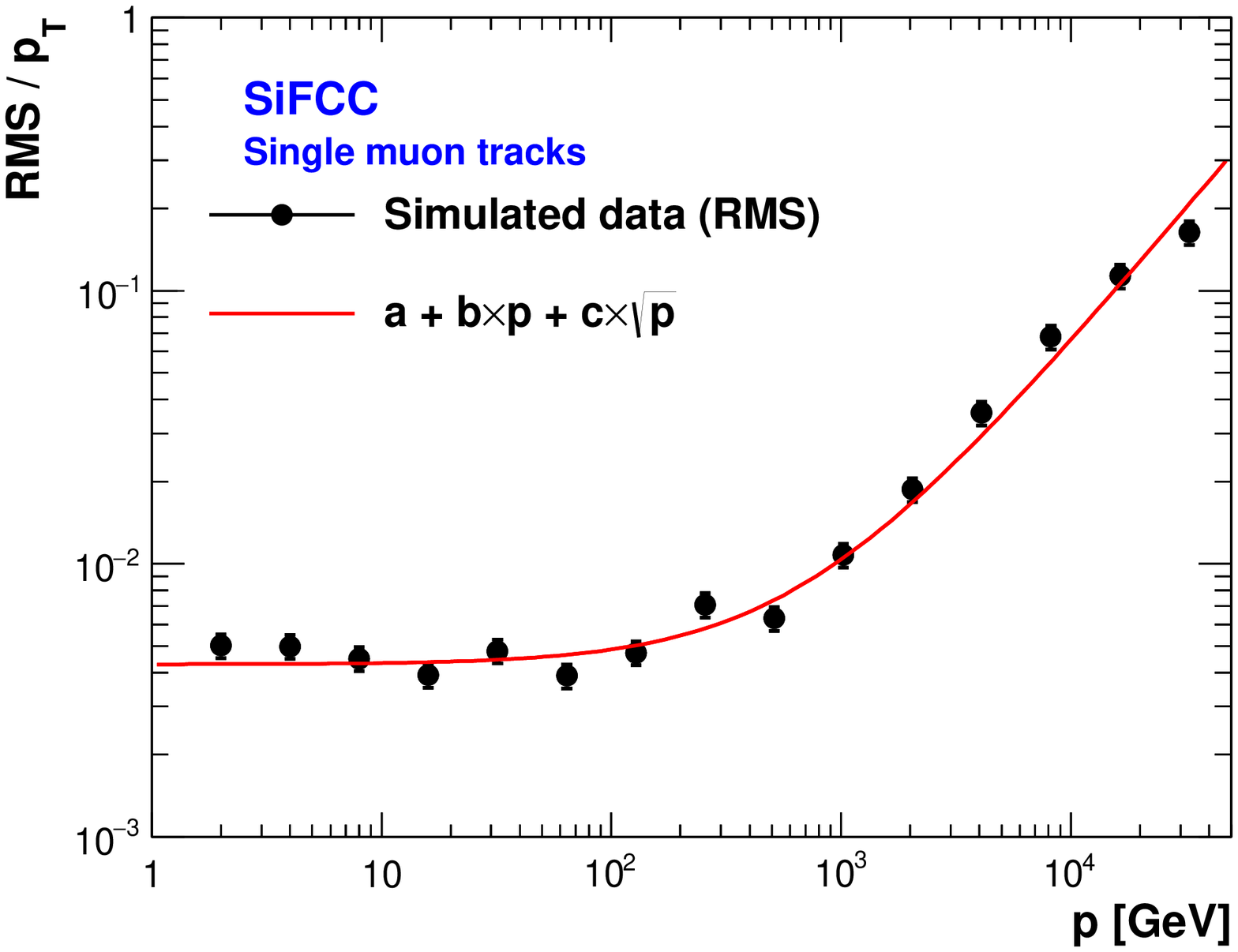}
  }
  \subfigure[$\sigma(p_T)/p_T$ vs $p_T$] {
  \includegraphics[width=0.46\textwidth]{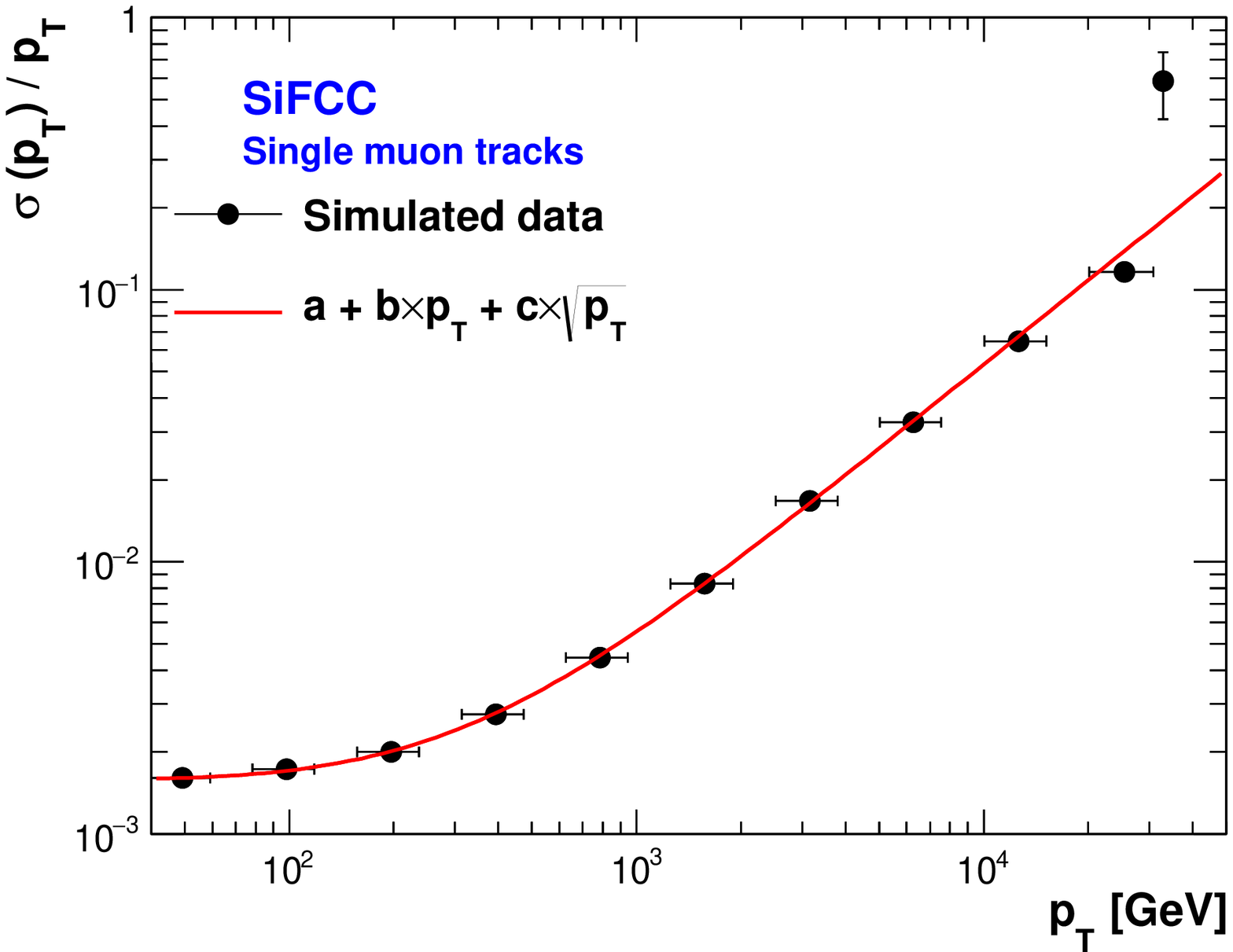}
  }
\caption{Fractional track $p_T$ resolution as a function of (a) $p$ and (b) $p_T$.
The resolution was found using the RMS of the ratio $p_T^{\rm reco}/p_T^{\rm true}$ for (a).
The $p_T$ resolution shown in (b)  
was determined using the width of a Gaussian fit for  $p_T^{\rm reco}/p_T^{\rm true}$. 
The figures show the fit functions used to describe the resolutions, with the fit parameters as given in the text.  
}
\label{fig:stracking}
\end{figure}

Figure~\ref{fig:stracking} shows the fractional track $p_T$ resolution $\sigma(p_T) / p_T$ estimated with single muons. 
At the highest momenta, $p = 16.4$~TeV and $p=32.8$~TeV,
the resolution is 10\%  and 20\%, respectively.
The resolution at the lowest momentum studied, $p=2$~GeV, is 0.5\%.

The simulated data are described by the following $p_T$ resolution function:
\begin{equation}
\frac{\sigma(p_T)}{p_T} = a + bp +  c \sqrt{p} 
\end{equation}
with the parameters $a=4.28\times 10^{-3}$, $b=6.23\times 10^{-6}$/GeV and $c=-4.73\times 10^{-6}$/$\sqrt{\rm GeV}$. 
The fit has a good $\chi^2$/ndf and is shown in Fig.~\ref{fig:stracking}(a).

Figure~\ref{fig:stracking}(b) shows the fractional track $p_T$ resolution $\sigma(p_T) / p_T$ as a function of $p_T$. 
  For this figure, 
 the distribution of the  ratio $p_T^{\rm reco}/p_T^{\rm true}$  was fit  using a  Gaussian distribution.
The width $\sigma$ of the Gaussian was used as a measure of the fractional $p_T$ resolution.  Also shown is a fit function which can be used
to parameterize the resolution  for fast simulations. The fit function is
\begin{equation}
\frac{\sigma(p_T)}{p_T} = a + b p_T + c \sqrt{p_T} 
\end{equation}
with the fit parameters $a=1.76\times 10^{-3}$, $b=5.77\times 10^{-6}$/GeV and $c=-6.31\times 10^{-5}$/$\sqrt{\rm GeV}$.

These results show that the tracking performance is within expectations for the tracking system with the given
readout segmentation and the solenoid magnetic field. Further optimizations are possible after 
event simulation and reconstruction of the most important physics channels \cite{Mangano:2016jyj,Contino:2016spe}.

\newpage
\section{Calorimeter performance for single particles}
\label{sec:calo}

The calorimeter response and resolution were studied by simulating single incident particles of different energies and species.
The particles were uniformly distributed in pseudorapidity and azimuthal angle.   
The energies of  these particles  
were reconstructed with the anti-$k_T$ jet algorithm~\cite{Cacciari:2008gp,Catani:1993hr} using calorimeter clusters as input. 
These clusters were built from calorimeter hits as explained in Sect.~\ref{sec:detector} after applying the corresponding sampling fractions. 
No other corrections were applied. The minimum transverse energy of the calorimeter clusters used to build jets was set to 400~MeV. 

The resolution and response were 
calculated by matching reconstructed particles with truth-level particles, and calculating the ratio $p_T^{\rm reco}/p_T^{\rm true}$.
The studies were carried out in the best understood central region of $|\eta|<1.5$, where the efficiency for measurements of low-momentum particles is highest.  
A  Gaussian fit was performed to the distribution of this ratio.
The response and resolution were calculated from the mean and the width $\sigma$ of the Gaussian fit, respectively. 
The resolution as a function of $p_T$ was described by the following fit function:
\begin{equation}
\frac{\sigma (p_T)}{p_T} = a / \sqrt{p_T} \oplus b, 
\label{resj}
\end{equation}
where $a$ is the sampling (stochastic) term, $b$ is the constant term and the symbol $\oplus$ denotes
a quadratic sum.

\begin{figure} 
\centering
  \subfigure {
  \includegraphics[width=0.46\textwidth]{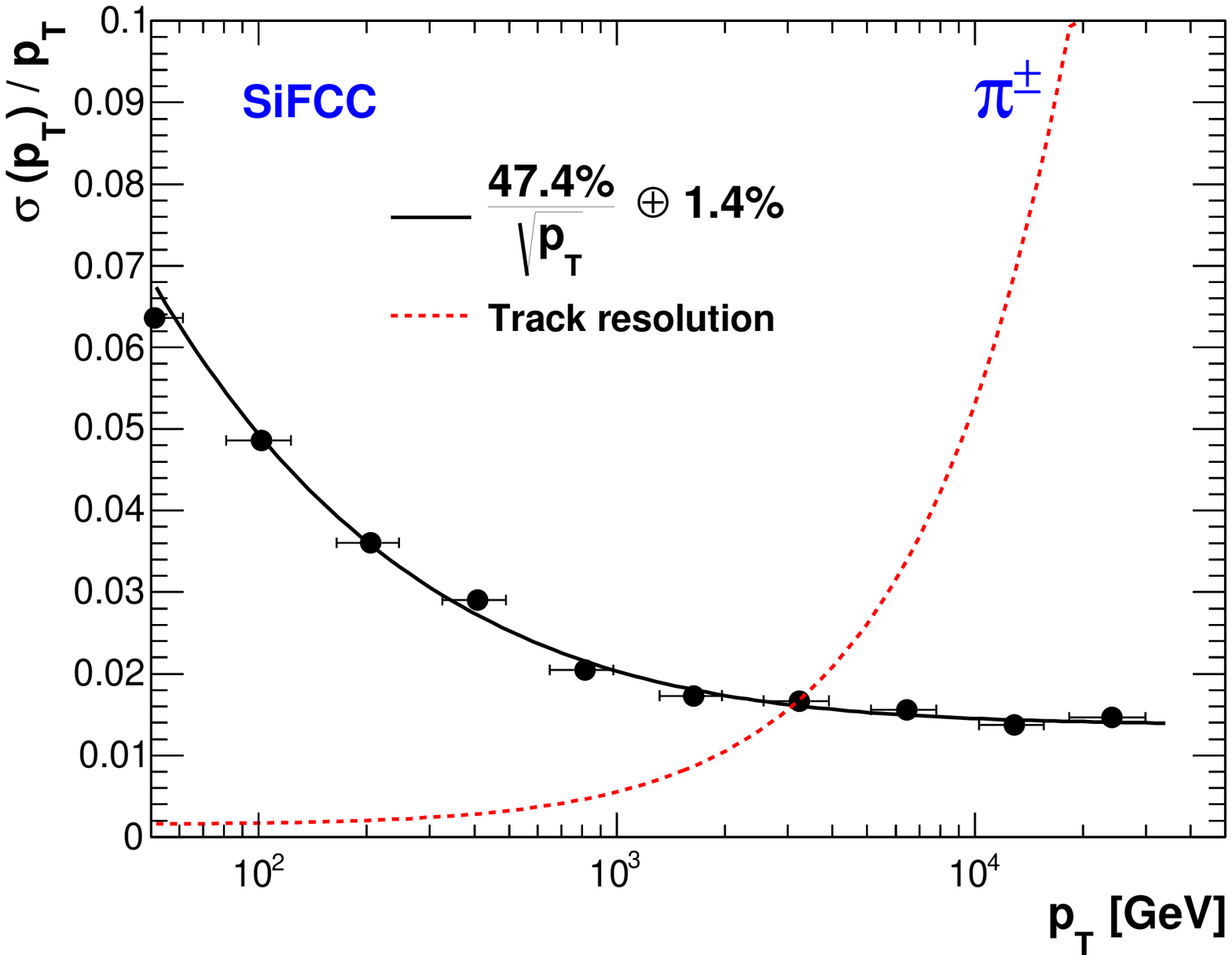}
  }
  \subfigure {
  \includegraphics[width=0.46\textwidth]{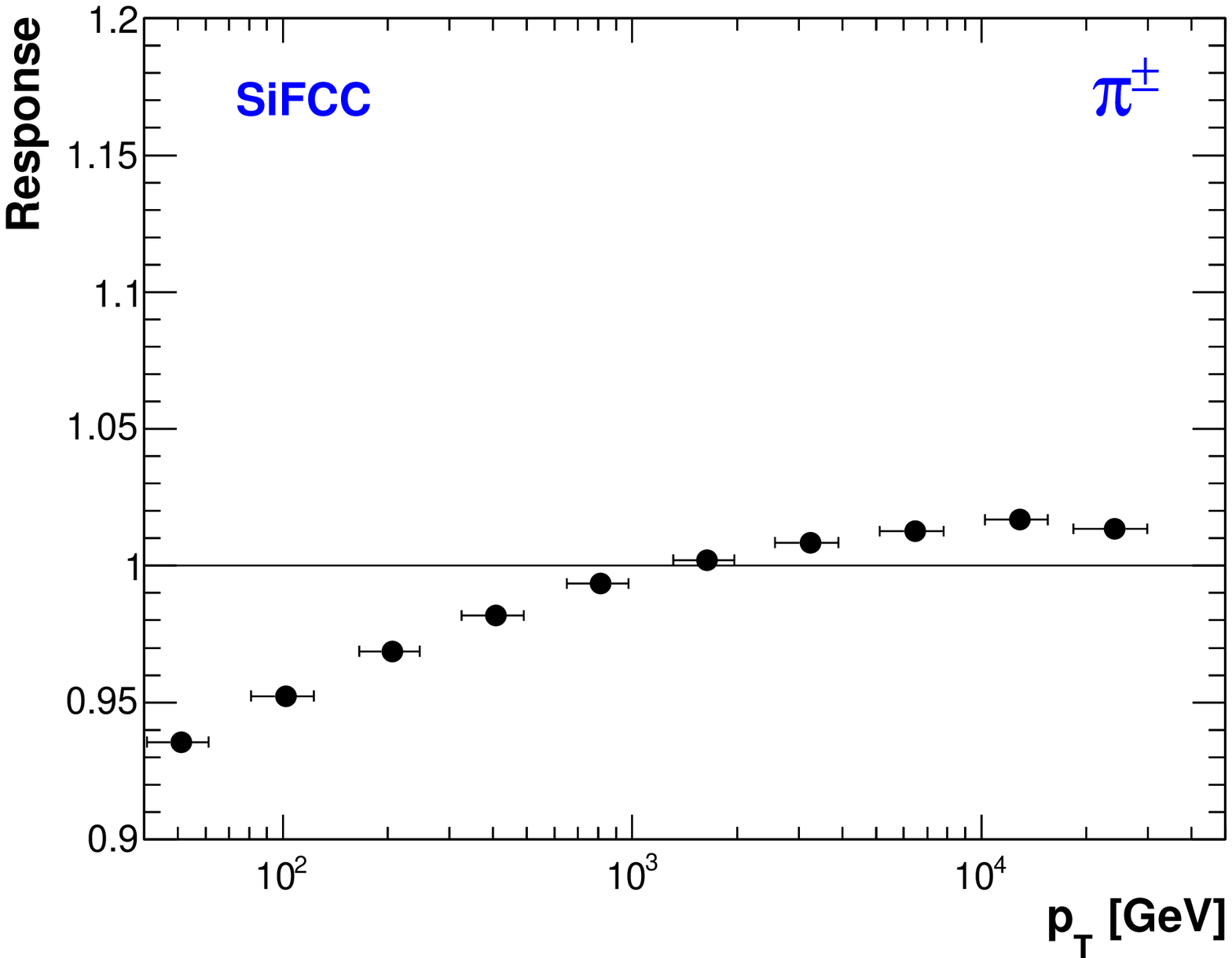}
  }
\caption{Calorimeter resolution (left) and response (right) to single $\pi^{\pm}$ in the pseudorapidity range $|\eta|<1.5$. 
 For comparison, the dotted line shows the track fractional $p_T$ resolution as discussed in Sect.~\ref{sec:tracking}. }
\label{fig:res:pion}
\end{figure}

\begin{figure}
\centering
  \subfigure {
  \includegraphics[width=0.46\textwidth]{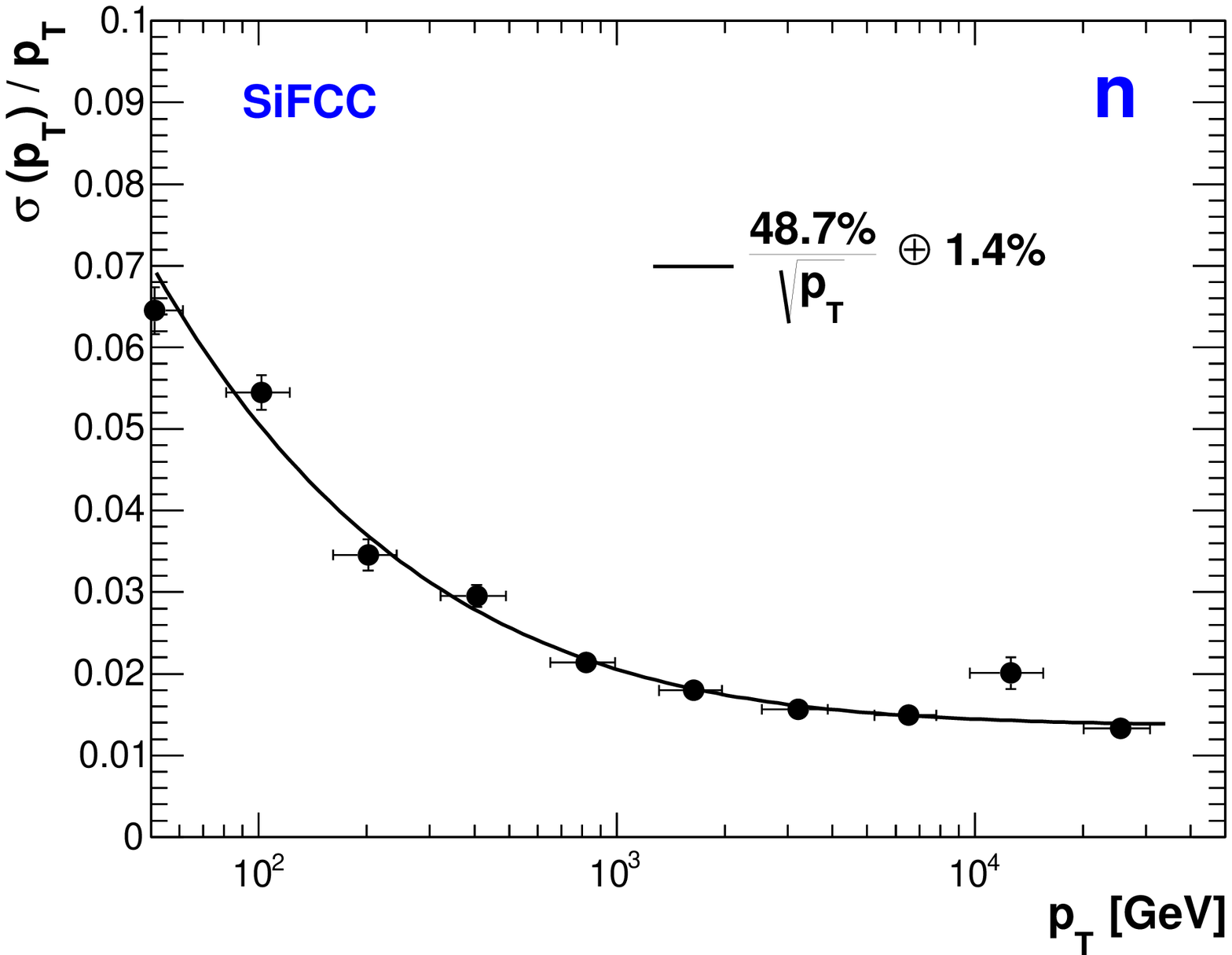}
  }
  \subfigure {
  \includegraphics[width=0.46\textwidth]{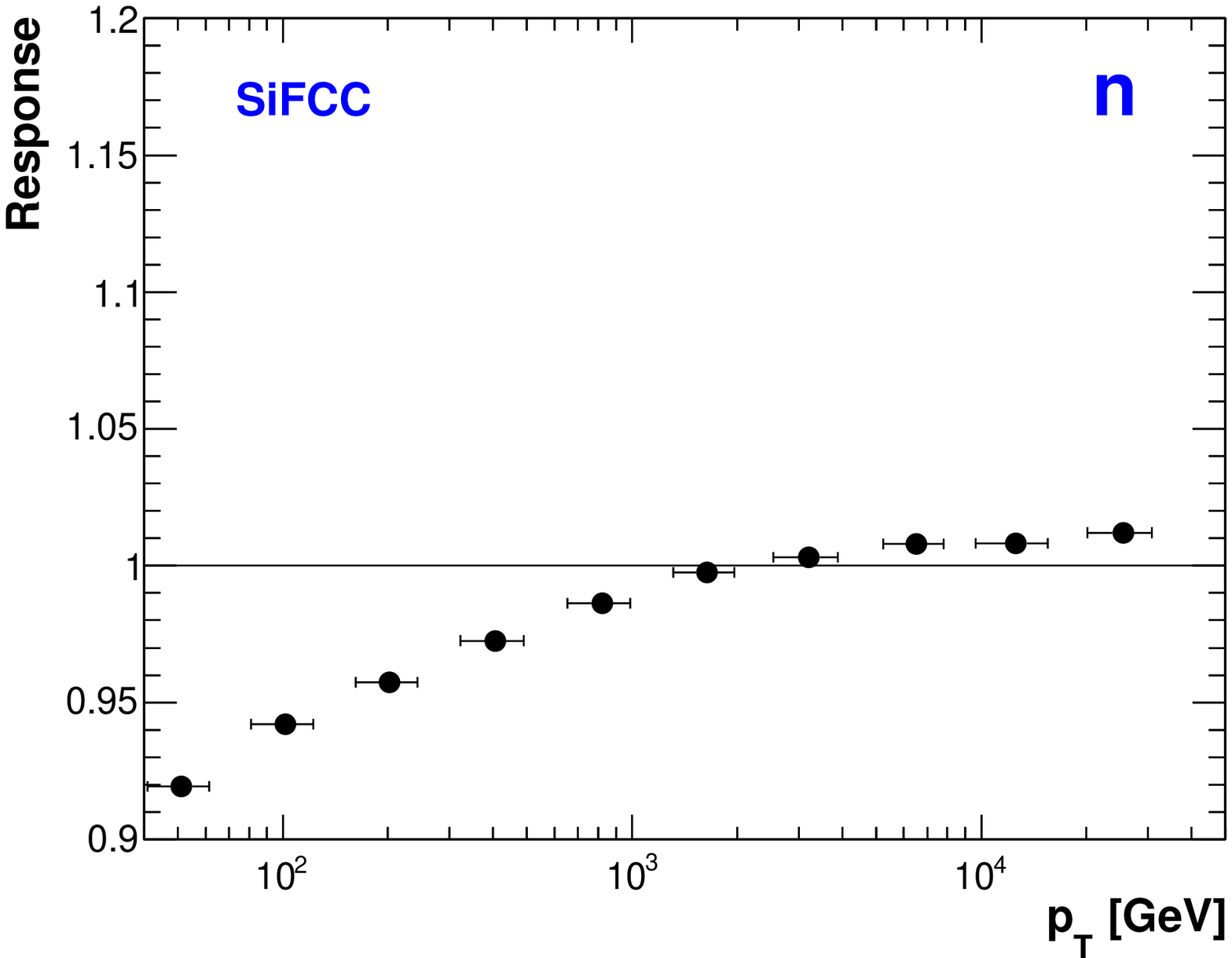}
  }

  \subfigure {
  \includegraphics[width=0.46\textwidth]{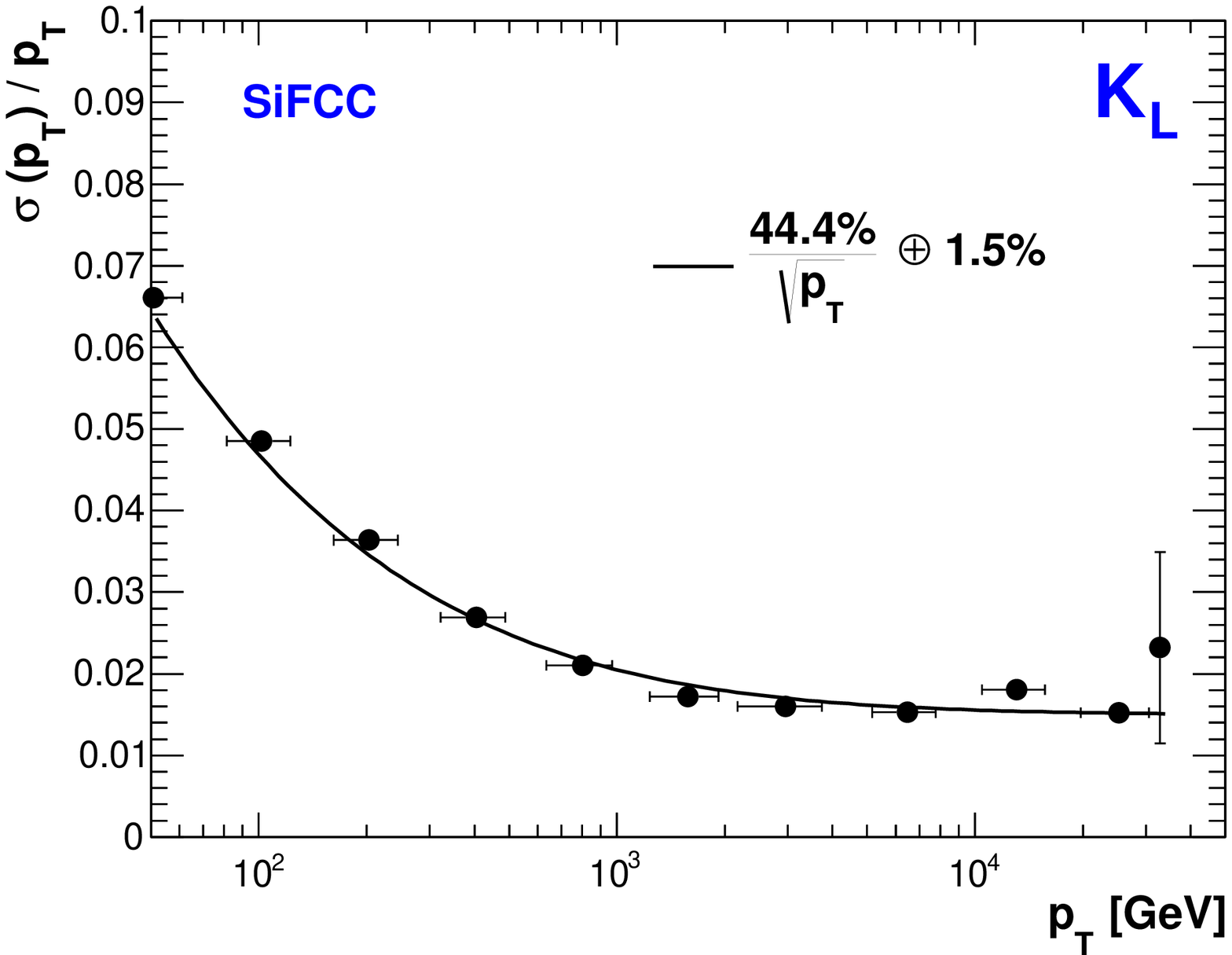}
  }
  \subfigure {
  \includegraphics[width=0.46\textwidth]{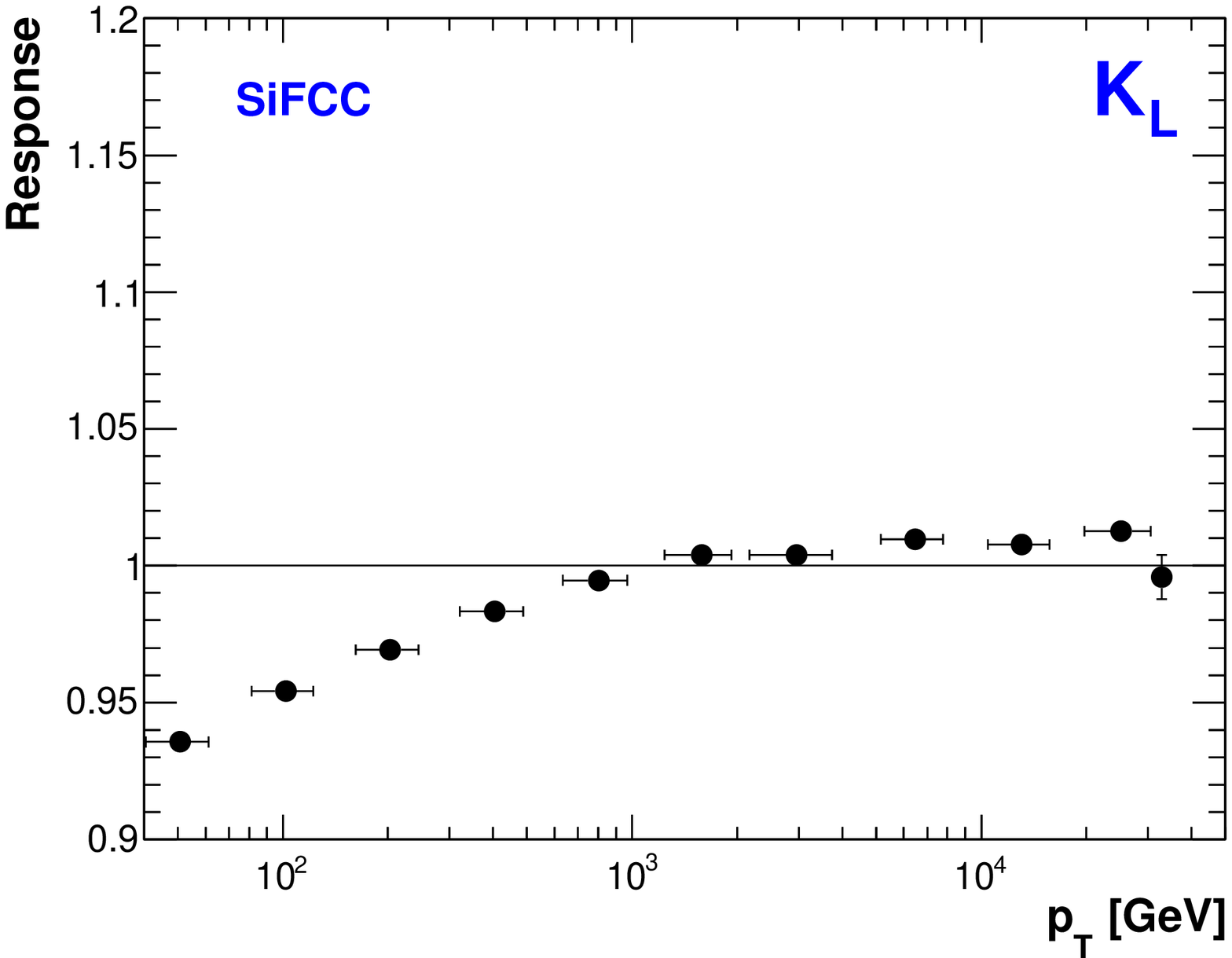}
  }
\caption{Calorimeter resolution (left) and response  (right) to single neutrons (top) and $K^0_L$ (bottom) in the pseudorapidity range $|\eta|<1.5$.}
\label{fig:res:klong}
\end{figure}

\begin{figure}
\centering
  \subfigure {
  \includegraphics[width=0.46\textwidth]{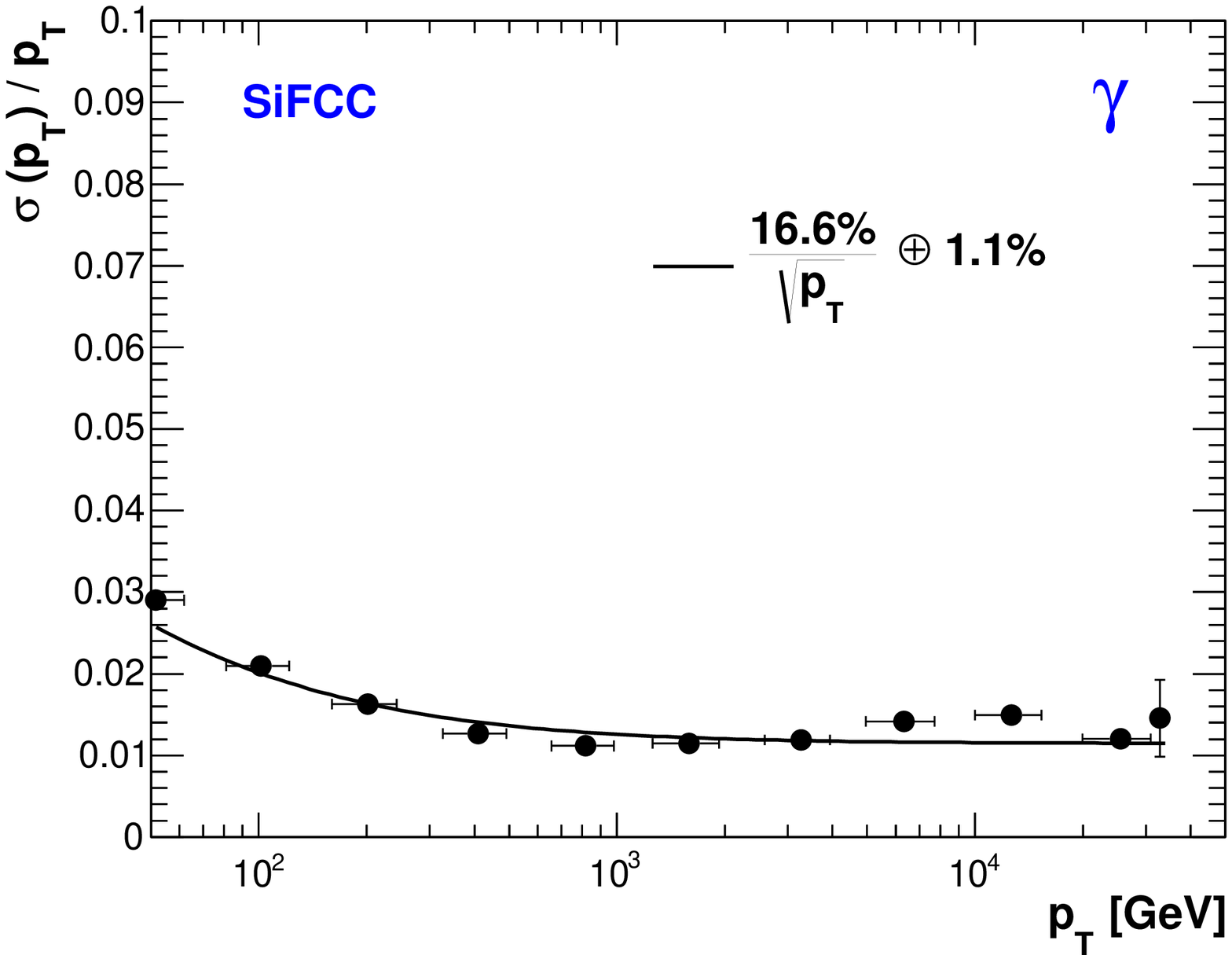}
  }
  \subfigure {
  \includegraphics[width=0.46\textwidth]{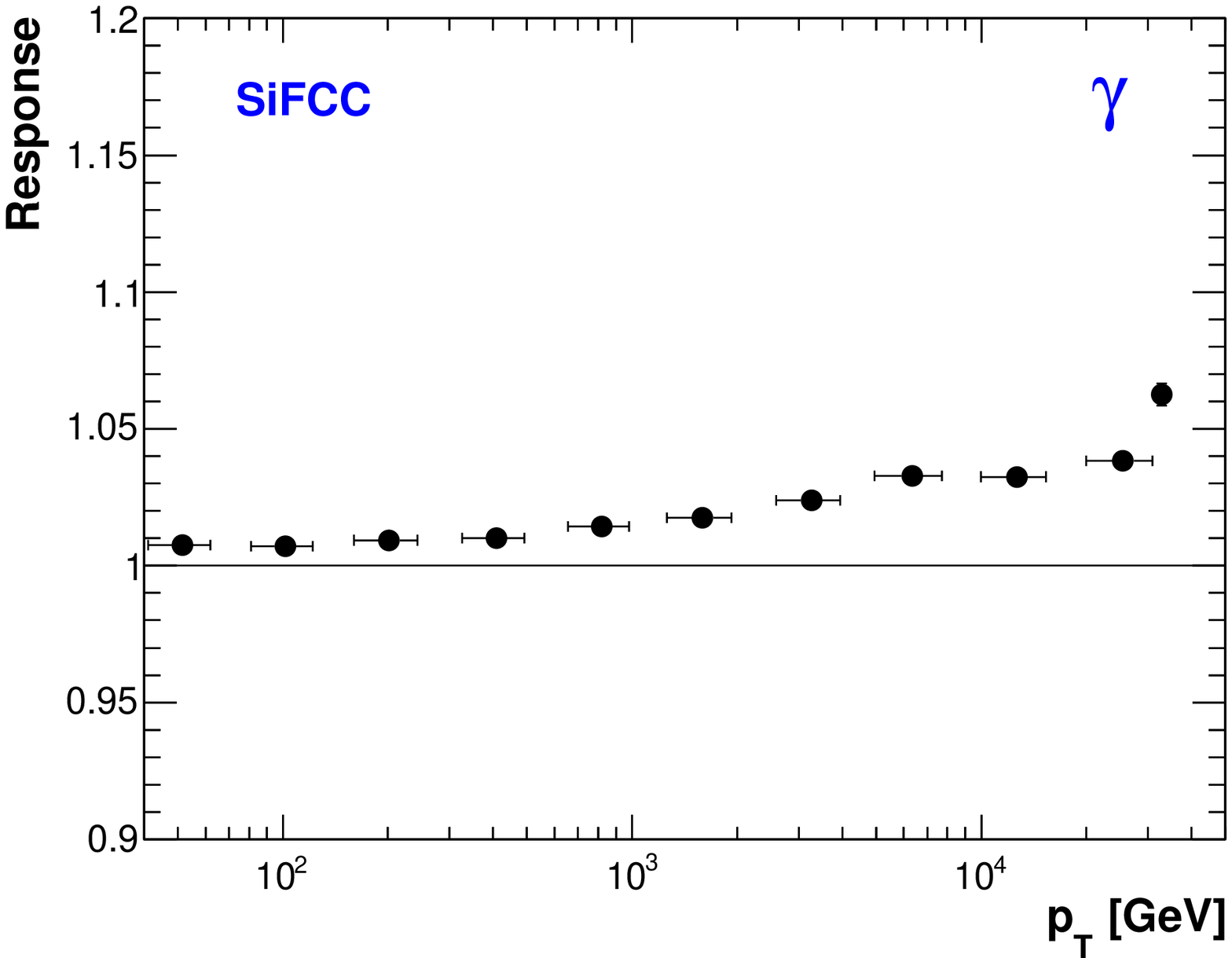}
  }
  \subfigure {
  \includegraphics[width=0.46\textwidth]{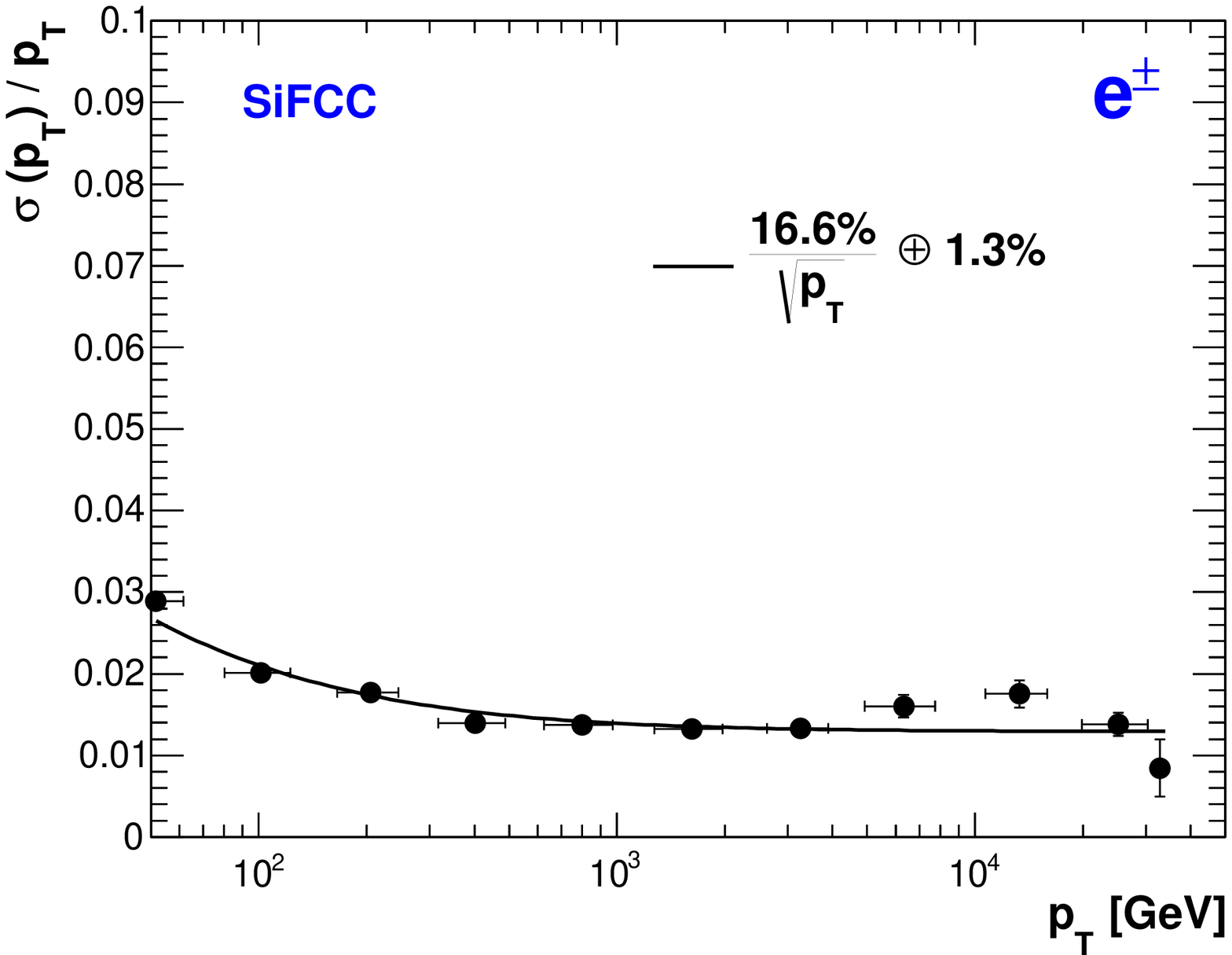}
  }
  \subfigure {
  \includegraphics[width=0.46\textwidth]{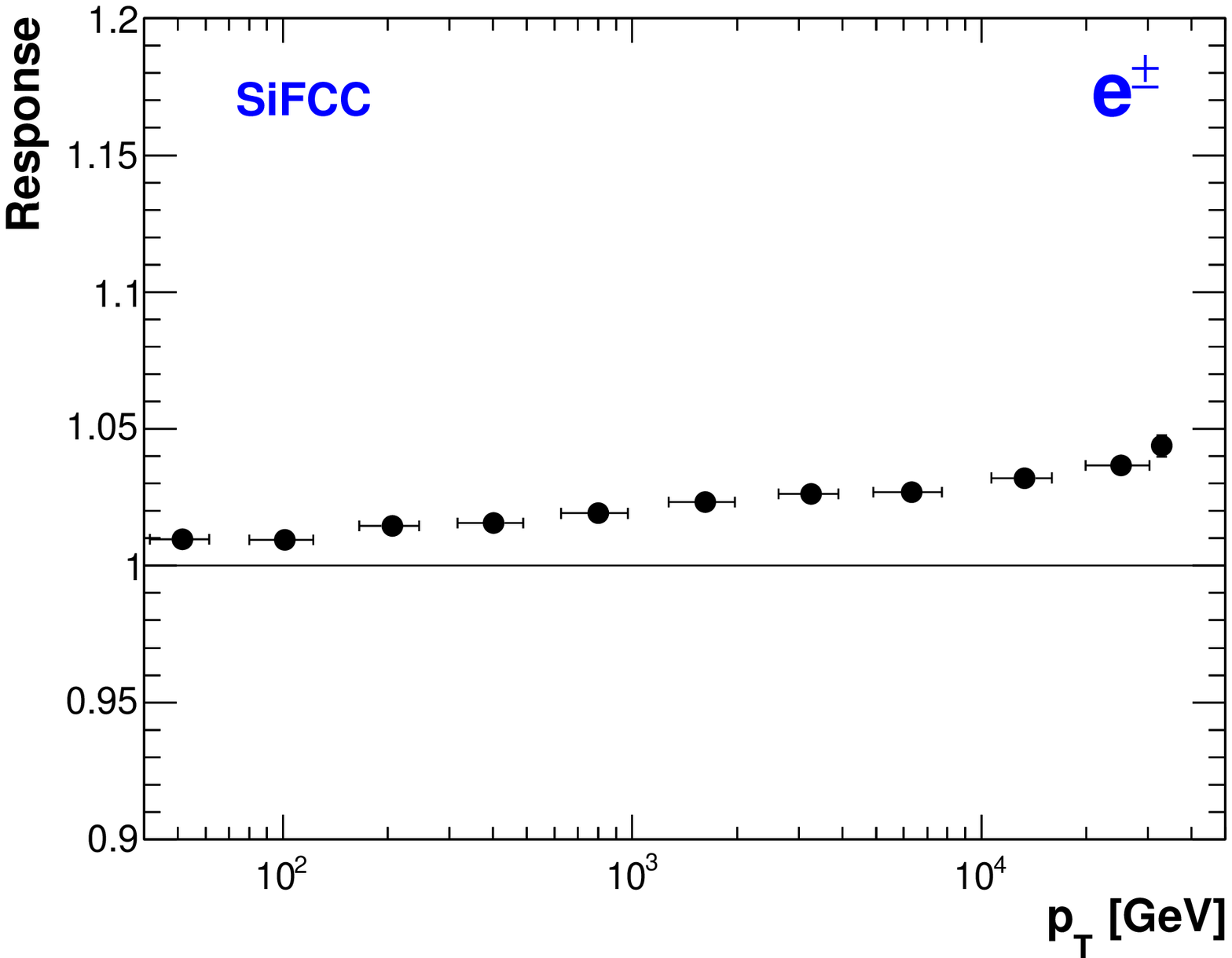}
  }
\caption{Calorimeter resolution (left) and response  (right) to single photons (top) and electrons / positrons  (bottom) in the pseudorapidity range $|\eta|<1.5$.}
\label{fig:res:elec}
\end{figure}

Figures~\ref{fig:res:pion}, \ref{fig:res:klong} and  \ref{fig:res:elec}
show the calorimeter resolution and response to single particles ($\pi^\pm$, $n$,  $K_L$, $\gamma$, $e^{\pm}$)  in the transverse
momentum range 2~GeV -- 32.8~TeV.
The sampling terms of the resolution function for  hadrons ($\pi^\pm$, $n$, $K_L$) is $44\% - 49\%$. 
For electrons and photons, it is close to $17\%$. 
The constant terms of the resolution function are in the range $1.1\% - 1.5\%$, which is somewhat smaller than for 
the studies of single pions using an alternative calorimeter setup~\cite{Carli:2016iuf} based on the ATLAS HCAL geometry.

For comparison with track-based measurements, Figure~\ref{fig:res:pion}
shows the tracking resolution function obtained in Sect.~\ref{sec:tracking}. 
The calorimeter measurement of particle momentum above $3$~TeV becomes more precise than the tracking measurement.

These studies show that the calorimeter response to single hadrons as a function of energy is non-linear, 
as expected for non-compensating ECAL and HCAL calorimeters with $e/h>1$. The response
 increases with energy, which is another expected effect for such calorimeters~\cite{Wigmans:2011zz}.  
The response to electromagnetic particles is almost linear. The results obtained with
a $\pi^0 \to \gamma \gamma$ source were consistent with the electron and photon results.

Our studies do not indicate a leakage out the back of  the hadronic calorimeter 
for single hadrons up to 32.8~TeV in transverse momentum.

\newpage
\section{Jet reconstruction}
\label{sec:jets}

Many physics channels of interest at a 100~TeV pp collider require 
good understanding of jet reconstruction in the transverse
momentum range from tens of GeV to tens of TeV.
For the jet performance study using the SiFCC detector setup, 
QCD dijet events for 100~TeV $pp$
collision energy were generated with the {\sc Pythia8}  Monte Carlo generator~\cite{Sjostrand:2006za}.
The MSTW2008LO68cl set of parton density functions \cite{Martin:2009iq} was used.
For jet clustering with truth-level particles as input, stable particles were selected
if their mean lifetimes are larger than $3\cdot 10^{-11}$ seconds.
Neutrinos were excluded from consideration in jet clustering.

The jets were reconstructed using the anti-$k_T$ jet algorithm 
\cite{Cacciari:2008gp,Catani:1993hr}.
The distance parameter of the  anti-$k_T$ jets was $R=0.4$ in order to make a meaningful
comparison with the calorimeter-based jets of the ATLAS experiment \cite{Aad:2012ag}. 
As in the case of single particles, jets were constructed from calorimeter clusters 
after correcting calorimeter hits by the sampling fractions. 
No other corrections were used, such as those related to the 
non-compensating calorimeters ($e/h>1$), out of the cone leakage, dead material and others.
As in the case of single-particle studies, the jet resolution and jet response were calculated in the 
pseudorapidity region of $|\eta(\rm{jet})|<1.5$.

\begin{figure}
\centering
\includegraphics[width=0.8\textwidth]{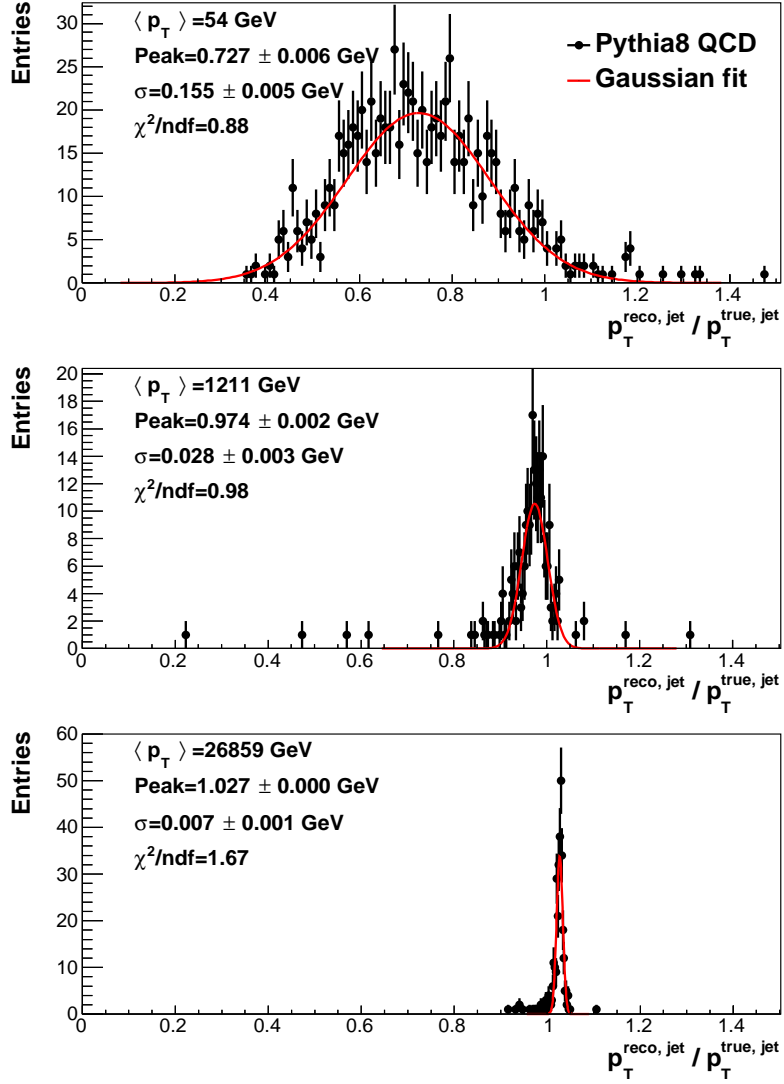}
\caption{The distributions of $p_T^{\rm reco,\>\rm{jet}}/p_T^{\rm true,\>\rm{jet}}$ 
for jets reconstructed with the anti-$k_T$ jet algorithm using the distance parameter of $0.4$.
The figure shows three ranges of the jet transverse momentum, together with Gaussian fits.
The mean values, the Gaussian widths and the value of $\chi^2/\rm{ndf}$ of the $\chi^2$ minimization procedure 
are indicated.} 
\label{fig:res_pt_3bin}
\end{figure}

Figure~\ref{fig:res_pt_3bin} shows the 
$p_T^{\rm reco,\>\rm{jet}}/p_T^{\rm true,\>\rm{jet}}$ distributions, where
$p_T^{\rm reco,\>\rm{jet}}$ is the reconstructed jet transverse momentum and $p_T^{\rm true,\>\rm{jet}}$ is the
jet transverse  momentum reconstructed from truth-level stable particles.  
Reconstructed jets and truth-level jets were matched within a distance of 
$0.2$ defined in azimuthal angle and pseudorapidity.
Figure~\ref{fig:res_pt_3bin} shows the $p_T^{\rm reco,\>\rm{jet}}/p_T^{\rm true,\>\rm{jet}}$ distributions 
in selected ranges of jet 
transverse momentum, including the lowest and highest $p_T^{\rm{jet}}$ 
studied in this paper. The distributions are well described by a Gaussian distribution.

\begin{figure}
\centering
  \subfigure[Jet resolution] {
  \includegraphics[width=0.43\textwidth]{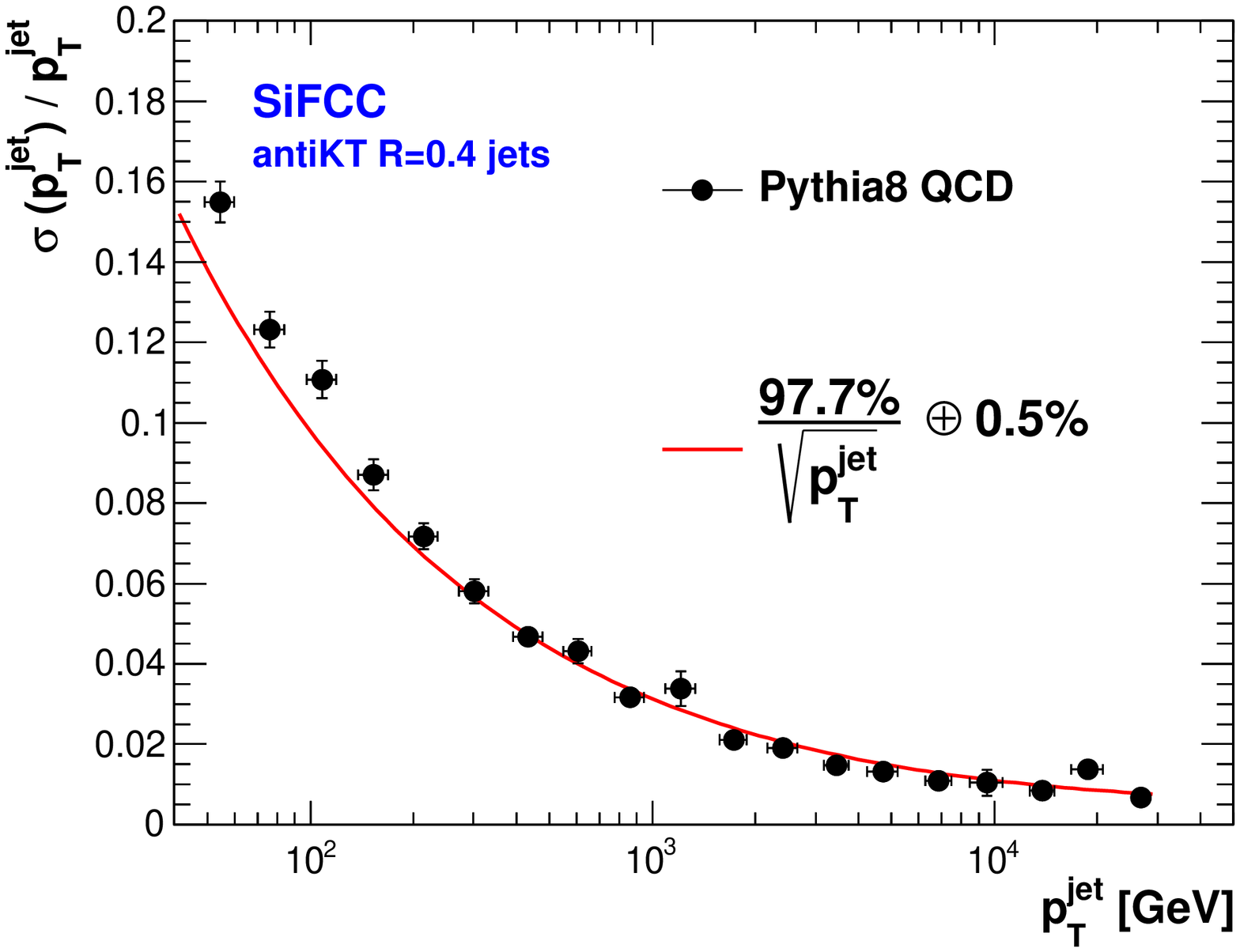}
  }
  \subfigure[Jet response] {
  \includegraphics[width=0.43\textwidth]{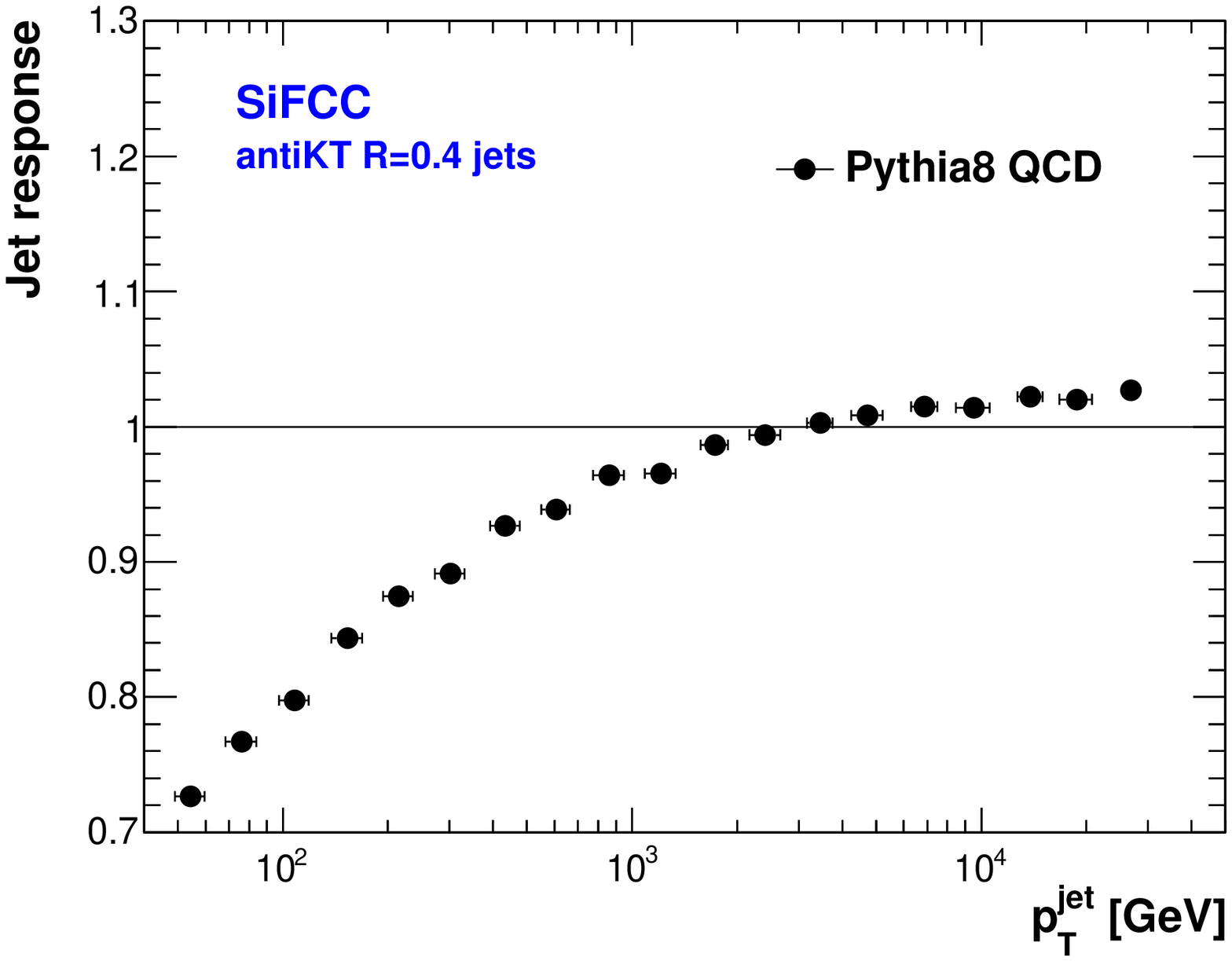}
  }
\caption{Resolution of anti-$k_T$ jets with the distance parameter $R=0.4$ as a function of jet transverse
momentum calculated using the Gaussian fit.  
The simulated data were fitted using the resolution function 
from 50~GeV to 26.4~TeV (shown with the red line). 
Figure (b) shows the jet response using the Gaussian fits for determination of the mean values
of the $p_T^{\rm reco,\> \rm{jet}}/p_T^{\rm true,\> \rm{jet}}$ distribution. The studies were performed using the {\sc Pythia8} Monte Carlo generator for $pp$ collision events at 100~TeV after simulation of the SiFCC detector response.} 
\label{fig:res:jets}
\end{figure}

The jet resolution as a function of the jet transverse momentum is shown in Figure~\ref{fig:res:jets}(a). 
It  was calculated using 
the width of the Gaussian fits illustrated in Fig.~\ref{fig:res_pt_3bin}.
The figure also shows the fit function Eq.~(\ref{resj}) with the sampling 
term $a/\sqrt{p_T^{\rm{jet}}}$ and a constant term. 
It should be noted that a noise term, which is  proportional to $1/p_T^{\rm{jet}}$, and which is frequently discussed
in the literature (see, for example, \cite{Aad:2012ag,Calkins:2013ega}), was not included into the fit since it was found that the current statistical precision 
cannot disentangle the noise term from the sampling term. 

Figure~\ref{fig:res:jets}(a) shows that the fit function discussed above gives a reasonable 
description of the jet resolution as a function of $p_T^{\rm{jet}}$
in the transverse-momentum region from 50~GeV to 26~TeV, with the value of $\chi^2/ndf\simeq 2.3$.
However, the fit function is below the simulations for  $p_T^{\rm{jet}}<0.2$~TeV. 
Alternative fits with the constant term fixed to a value in the range $1\%-2\%$, or with a reduced 
$p_T^{\rm{jet}}$ range, give similar values of $\chi^2/ndf$. 
The jet response shown in Figure~\ref{fig:res:jets}(b)  increases as a function of $p_T^{\rm{jet}}$,
and approaches unity for large transverse momenta, similar to the single-particle responses 
shown in Figs.~\ref{fig:res:pion}, \ref{fig:res:klong} and \ref{fig:res:elec}.

The results presented in  Fig.~\ref{fig:res:jets}  indicate a promising quality of 
jet reconstruction by the SiFCC detector, with
the estimated jet resolution
similar to a typical resolution for calorimeter-based jets of the ATLAS experiment
in the region below 1~TeV. For example,  
the fractional jet resolution of anti-$k_T$ jets with $R=0.4$  for the ATLAS experiment \cite{Aad:2012ag} 
is about $0.1$ for $p_T^{\rm{jet}}\simeq 100$~GeV  
and $0.05$ for jets  with $p_T^{\rm{jet}}\simeq 0.5$~TeV, which is similar to the
resolution shown in Fig.~\ref{fig:res:jets}.

These initial studies provide a first glimpse of  jet reconstruction in the energy range of tens of TeV.
The main conclusion of this study is that 
the constant term below $1-2\%$ is achievable  
for calorimeter-based jets with transverse momentum above $26$~TeV  
using the detector setup discussed in this paper.
A significant improvement for the jet-energy resolution at transverse momentum 
below 0.5~TeV is expected for the PFA approach.

\newpage
\section{Double-particle studies}

A key consideration for hadronic calorimeters for future collider detectors  is the impact of granularity on resolving 
energy deposits from pileup vertices and  highly-boosted jet topologies.
This understanding will be extremely important for measuring jet substructure for highly-boosted heavy objects such as $W$, $Z$ and $H$ bosons and top quarks.  
 Hadronic calorimeters currently in use have cell sizes which are larger  than the nuclear interaction length, $\lambda_I$.
In this section, the use of smaller cell sizes to resolve individual hadrons is investigated. 

To study the effect of calorimeter granularity on  hadron reconstruction, the density of energy reconstructed in calorimeter cells was investigated 
for events with pairs of $K^0_L$ produced at various angular separations. 
Since these neutral 
hadrons provide no tracking information, their position and energy measurements can only be obtained from calorimetry. 
Double-$K^0_L$ samples are simulated 
with energies of 100~GeV or 1000~GeV for both particles.
Each event has two particles separated by an azimuthal angle, $\Delta \phi^{K}$, keeping $\eta^{K} = 0$ for both particles.
The particles are separated  in multiples of $0.5^{\circ}$ (about 0.009~rad),  starting from $\Delta \phi^{K} = 0$,  and stepping them apart 
until they have the separation of $10^{\circ}$.  

\begin{figure}
\centering
  \subfigure[$20 \times 20$~cm HCAL cells and $2 \times 2$~cm ECAL cells] {
  \includegraphics[width=0.45\textwidth]{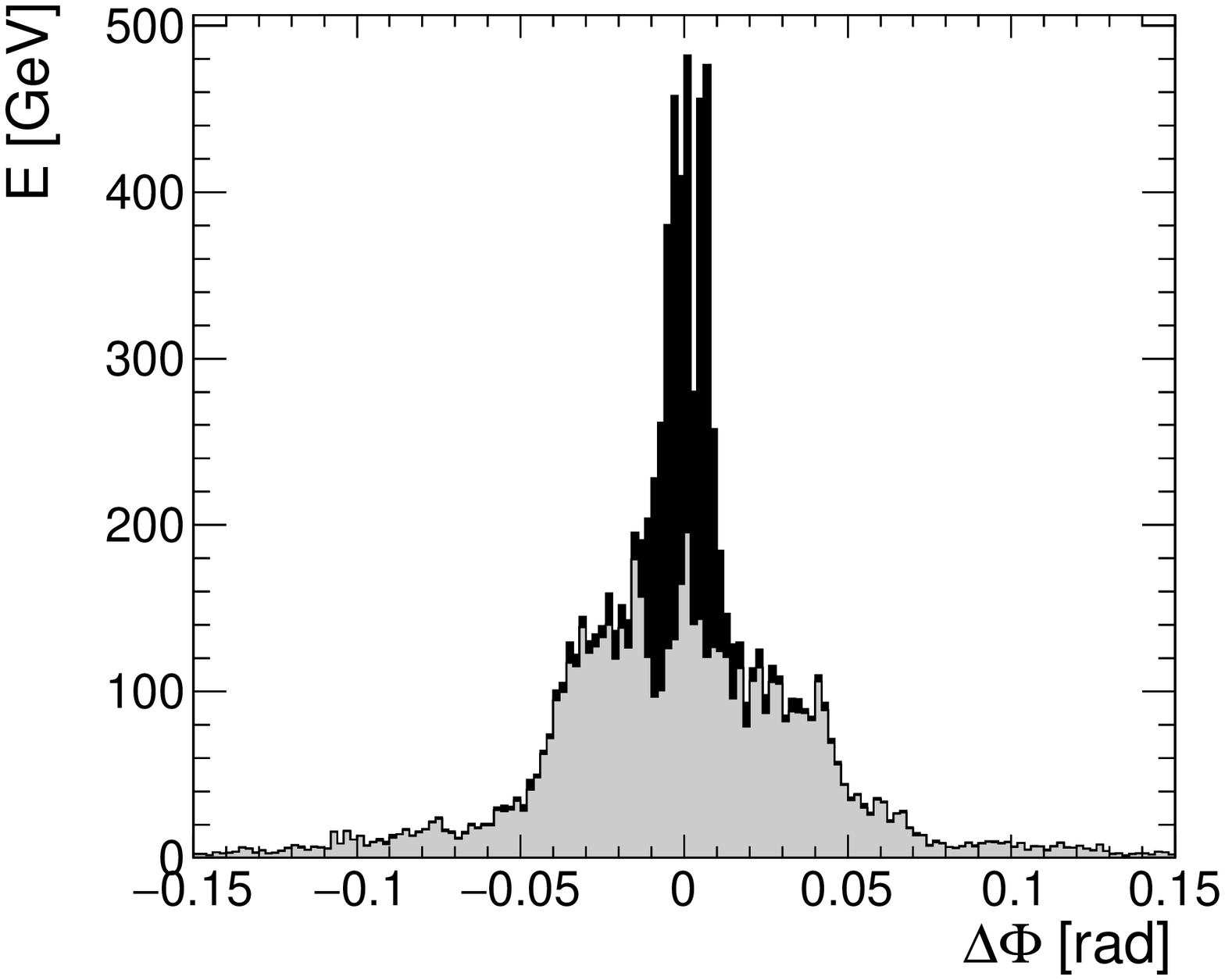}
  \includegraphics[width=0.45\textwidth]{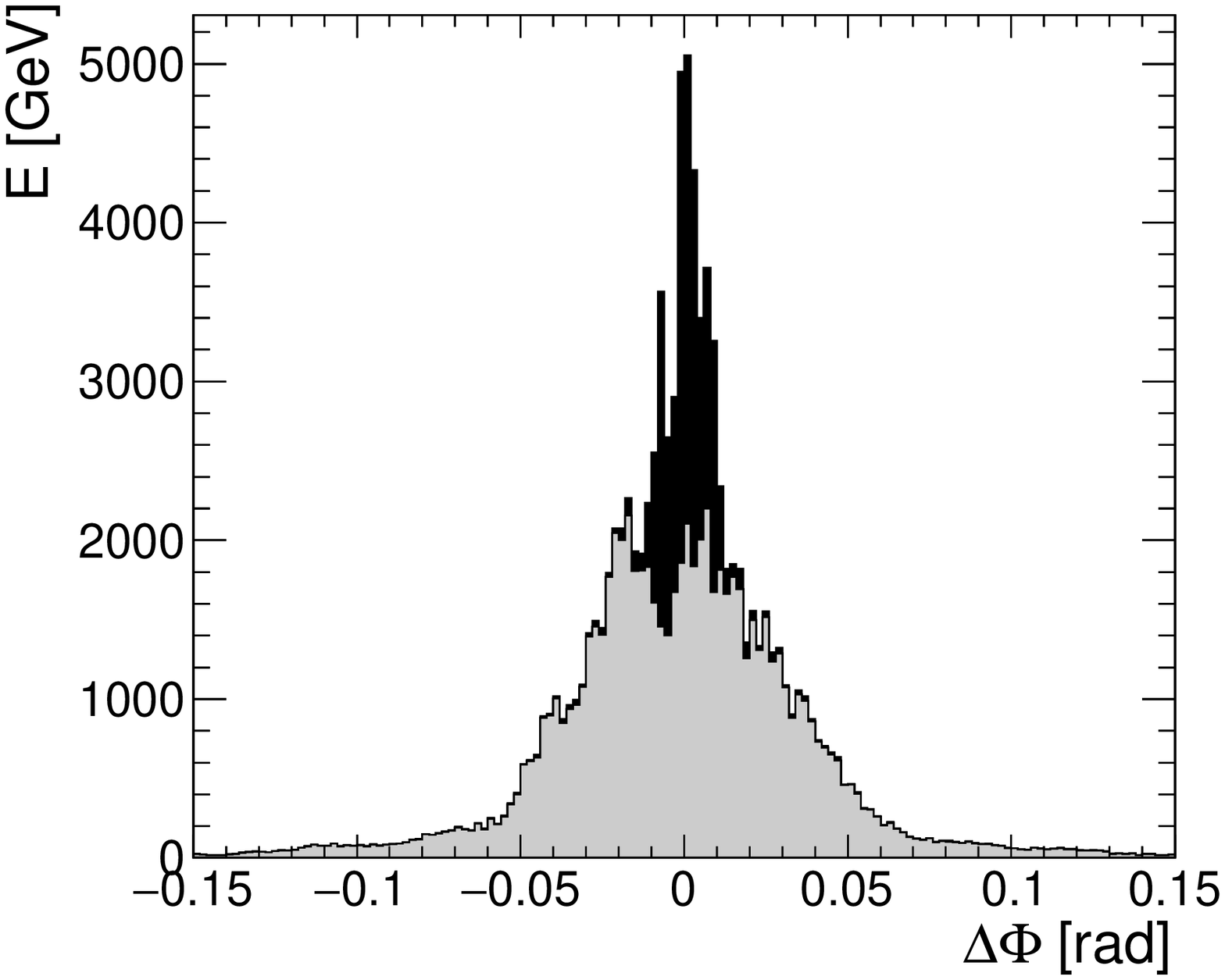}
  }
  \subfigure[$5 \times 5$~cm HCAL cells and $2 \times 2$~cm ECAL cells] {
  \includegraphics[width=0.45\textwidth]{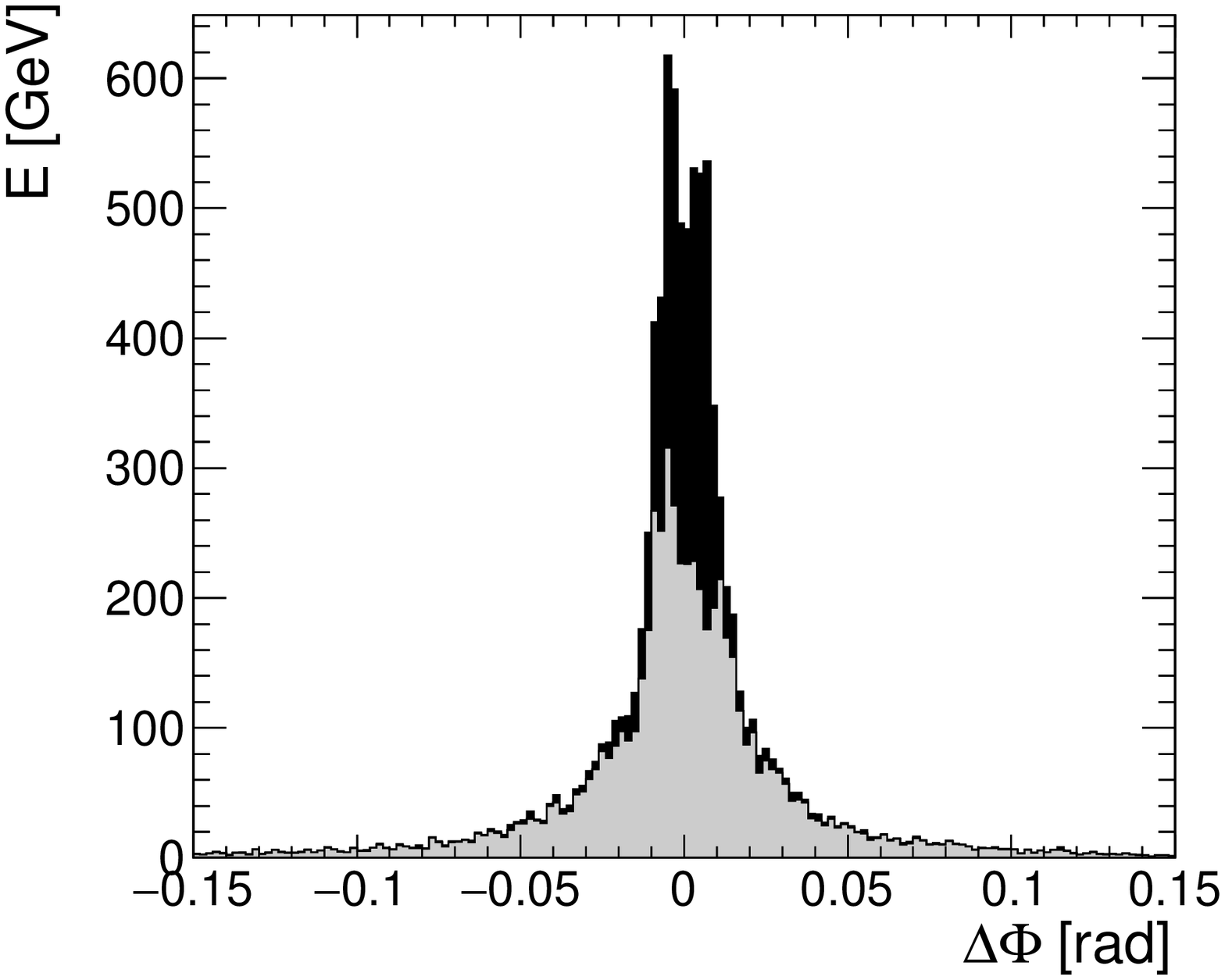}
  \includegraphics[width=0.45\textwidth]{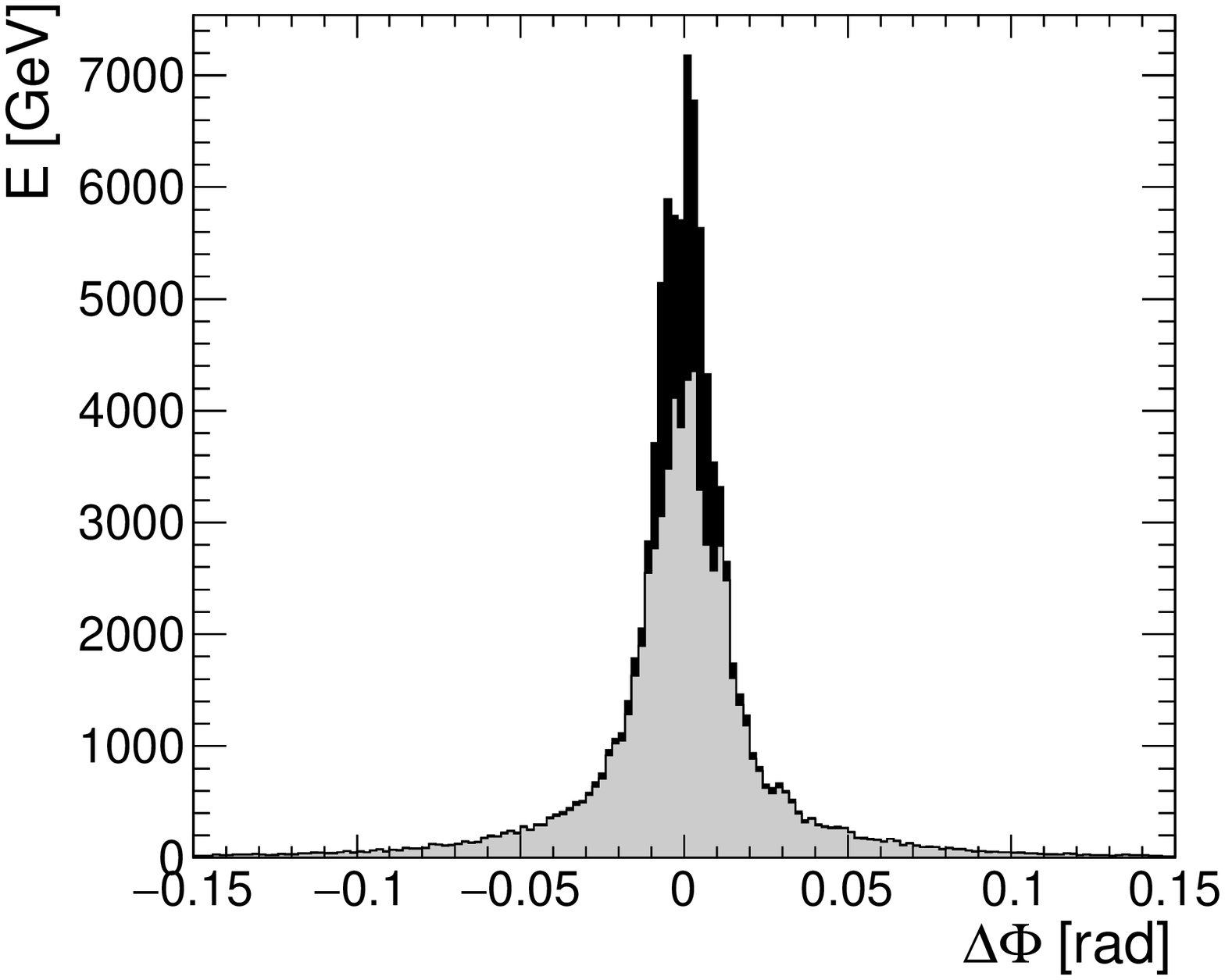}
  }
  \subfigure[$1 \times 1$~cm HCAL cells  and $3 \times 3$~mm ECAL cells] {
   \includegraphics[width=0.45\textwidth]{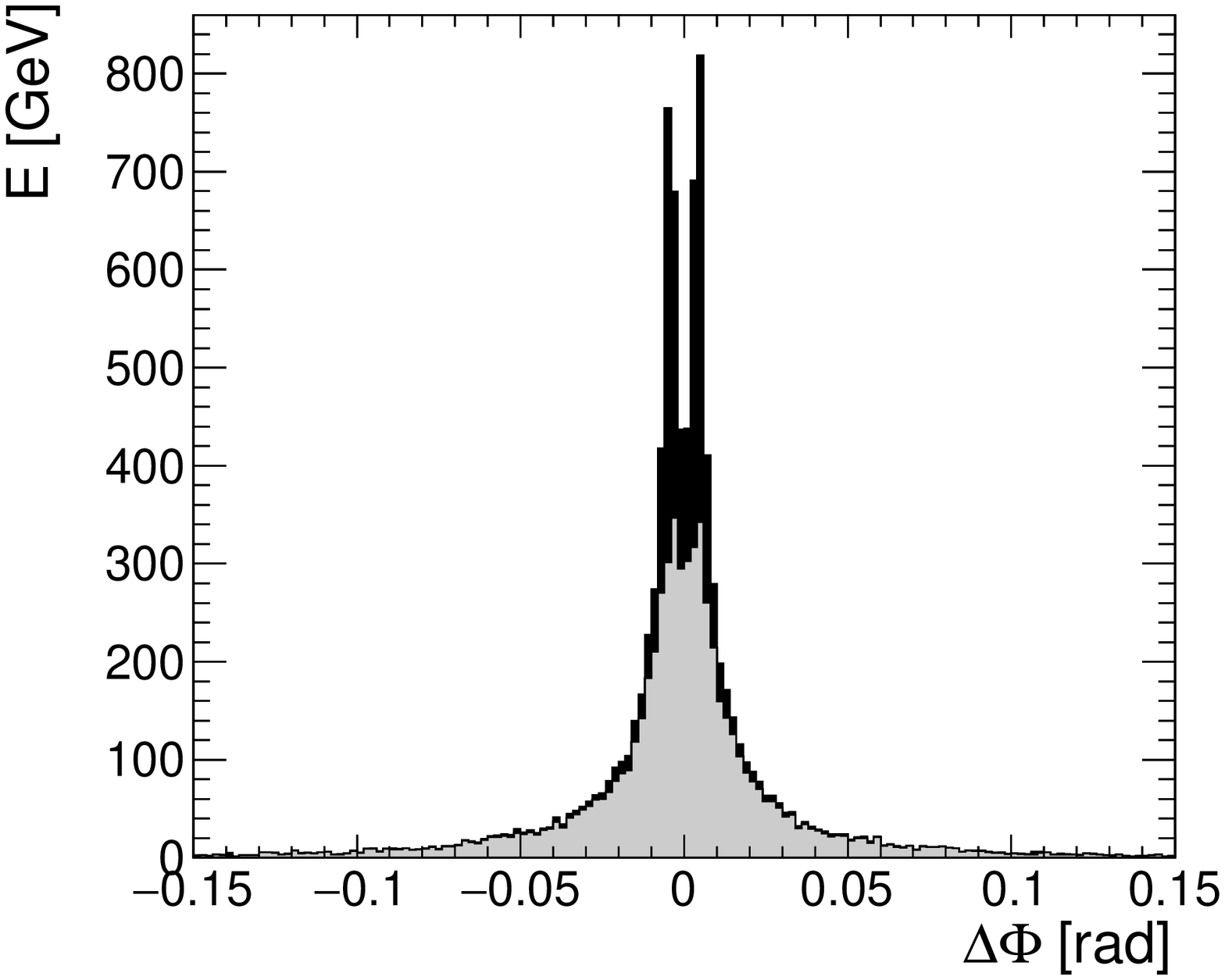}
   \includegraphics[width=0.45\textwidth]{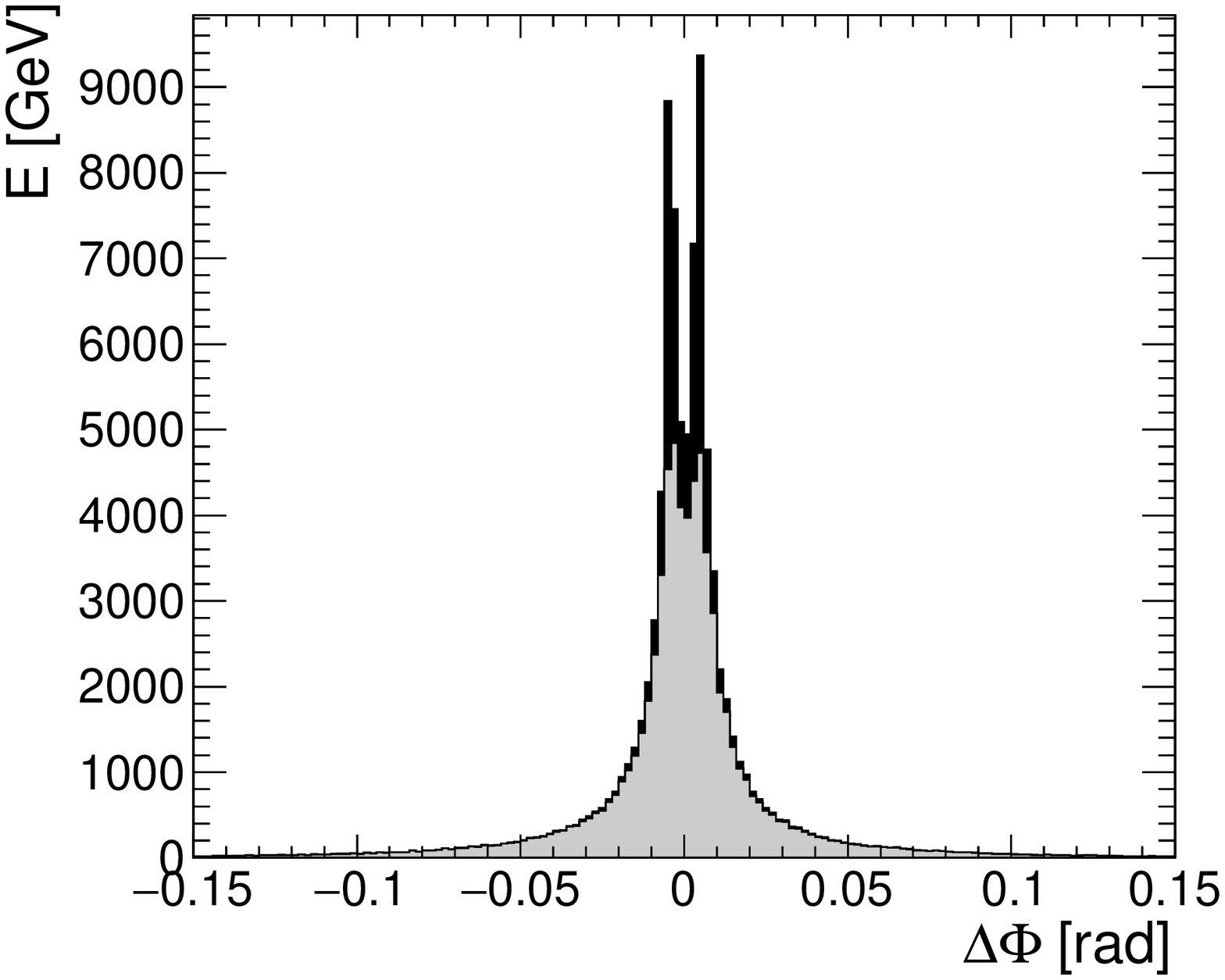}
 }
\caption{Azimuthal distribution of energy deposition for pair of incident $K^0_L$ particles at 100~GeV (left) and 1000~GeV (right), with the angular separation of 
$\Delta \phi^{K}=0.009$~rad.
Electromagnetic calorimeter cells are indicated in black while hadronic calorimeter  cells are indicated in gray.}
\label{fig:doublek1}
\end{figure}

\begin{figure} 
\centering
  \subfigure[$20 \times 20$~cm HCAL cells  and $2 \times 2$~cm ECAL cells] {
  \includegraphics[width=0.45\textwidth]{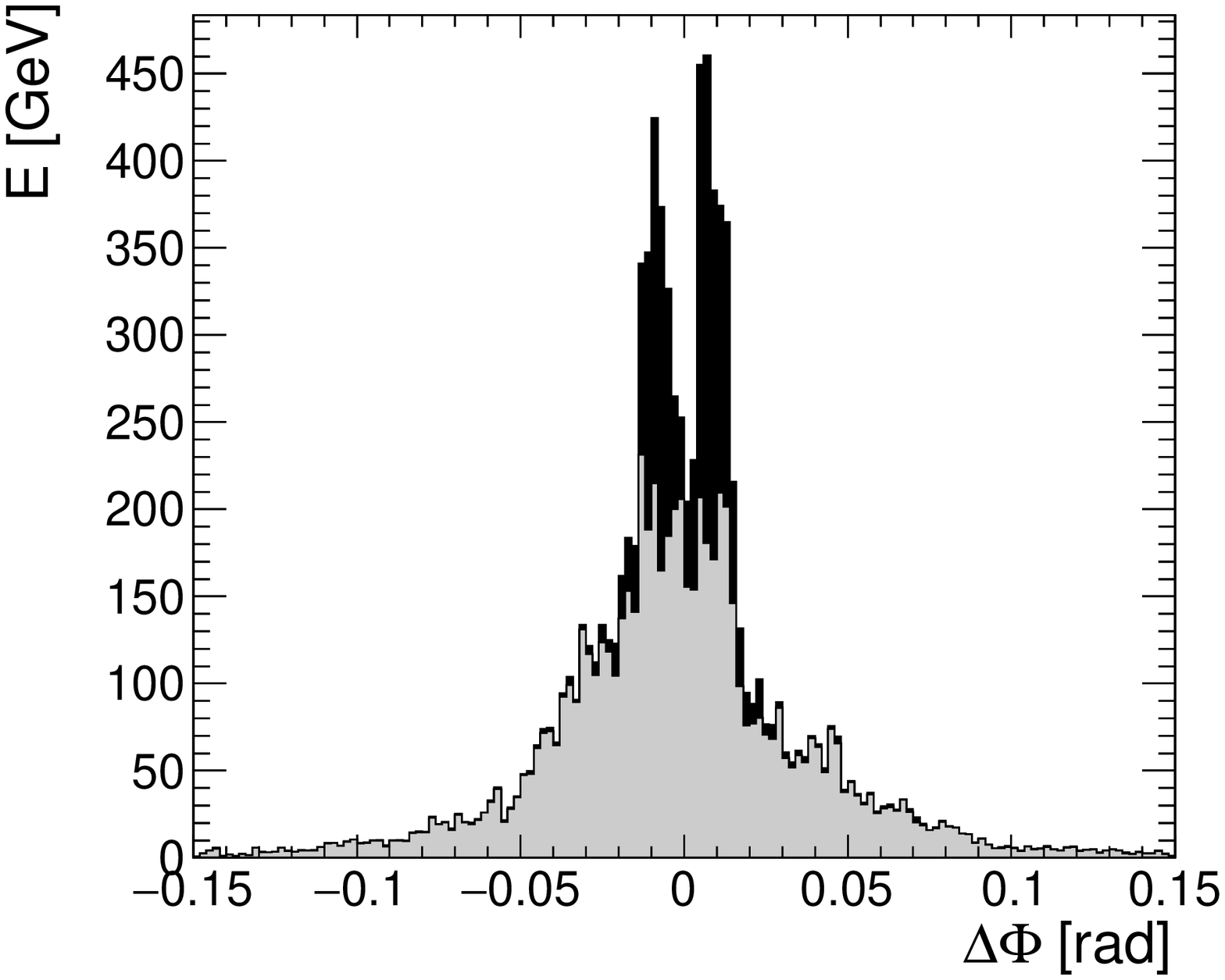}
  \includegraphics[width=0.45\textwidth]{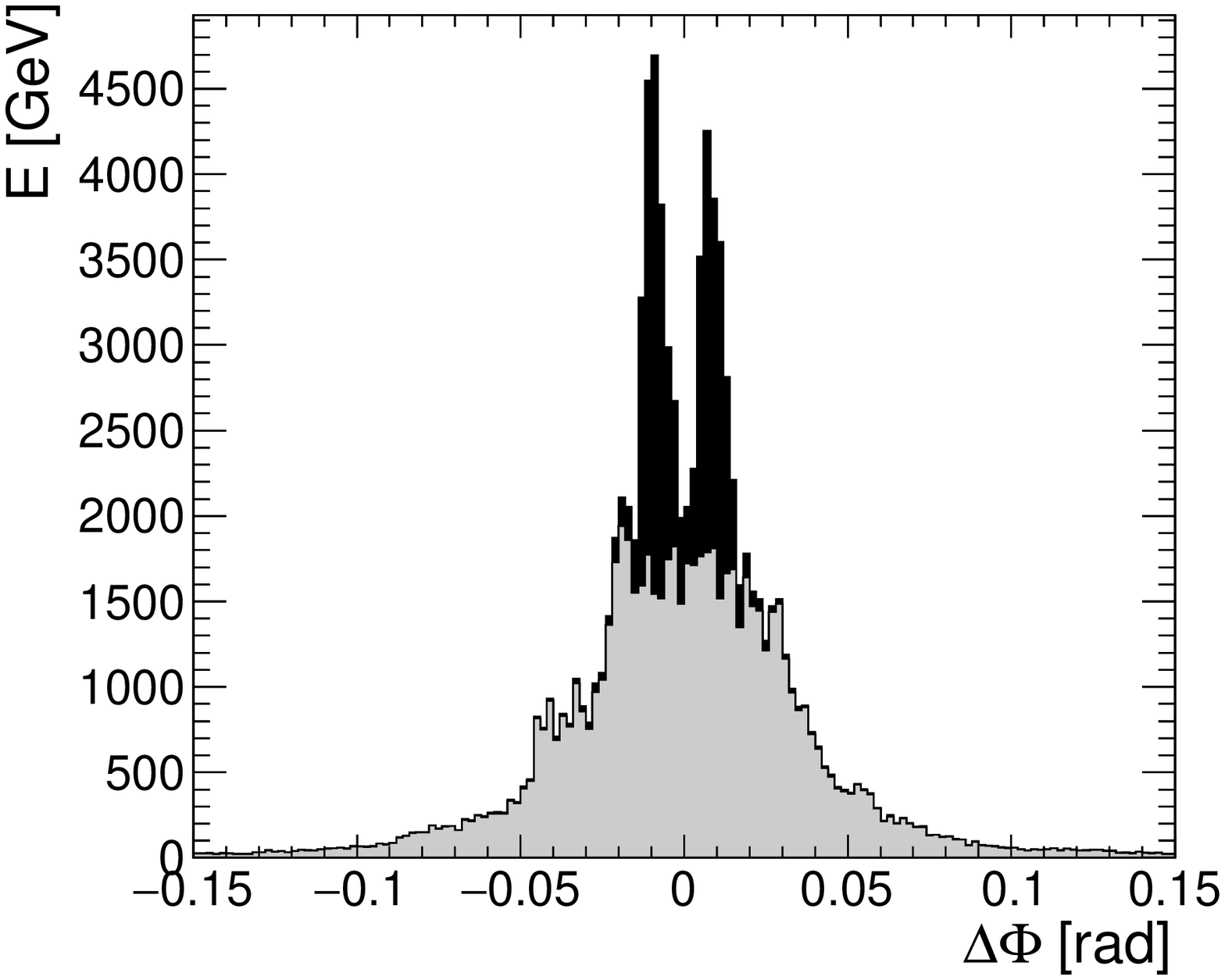}
  }
  \subfigure[$5 \times 5$~cm HCAL cells and $2 \times 2$~cm ECAL cells] {
  \includegraphics[width=0.45\textwidth]{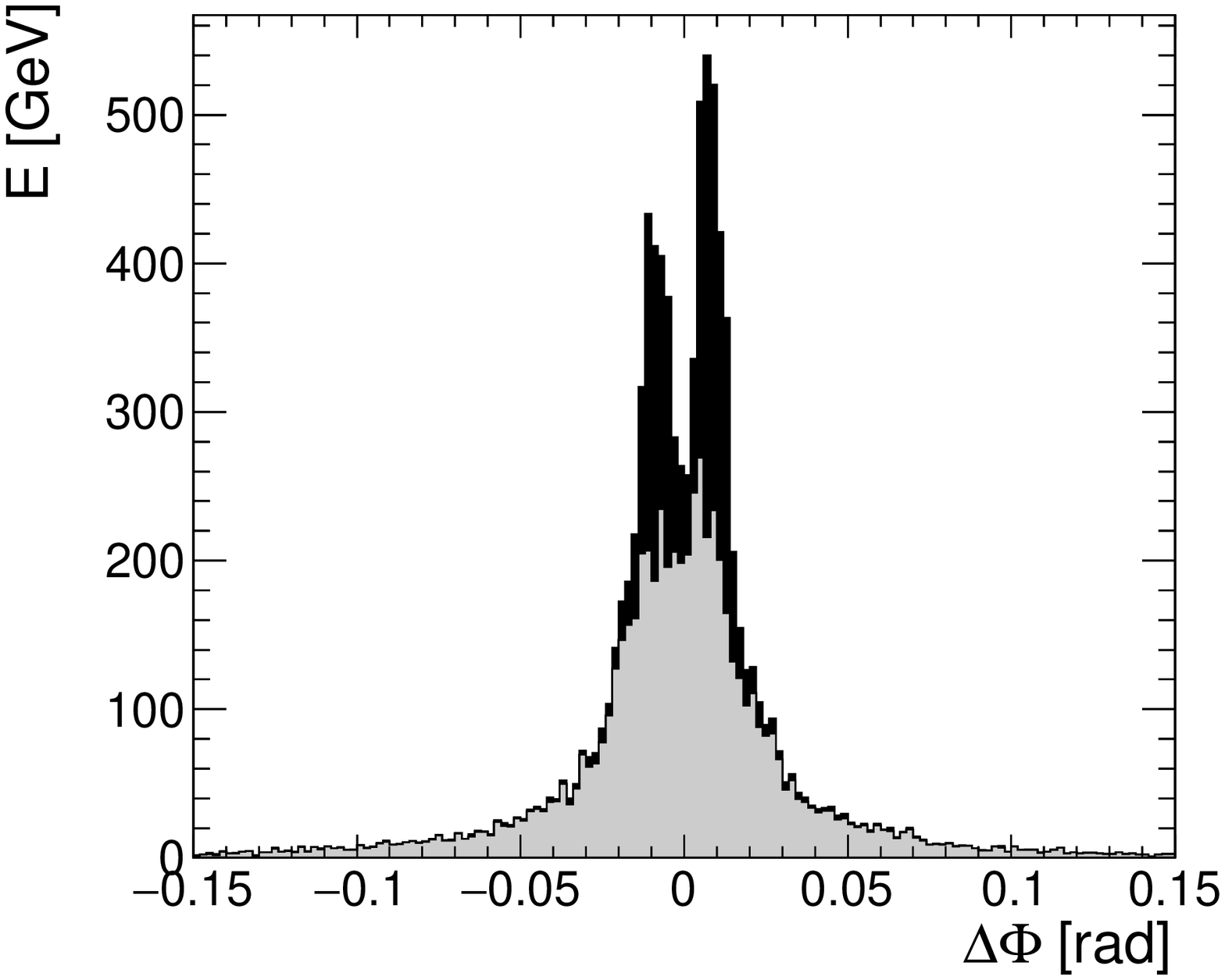}
  \includegraphics[width=0.45\textwidth]{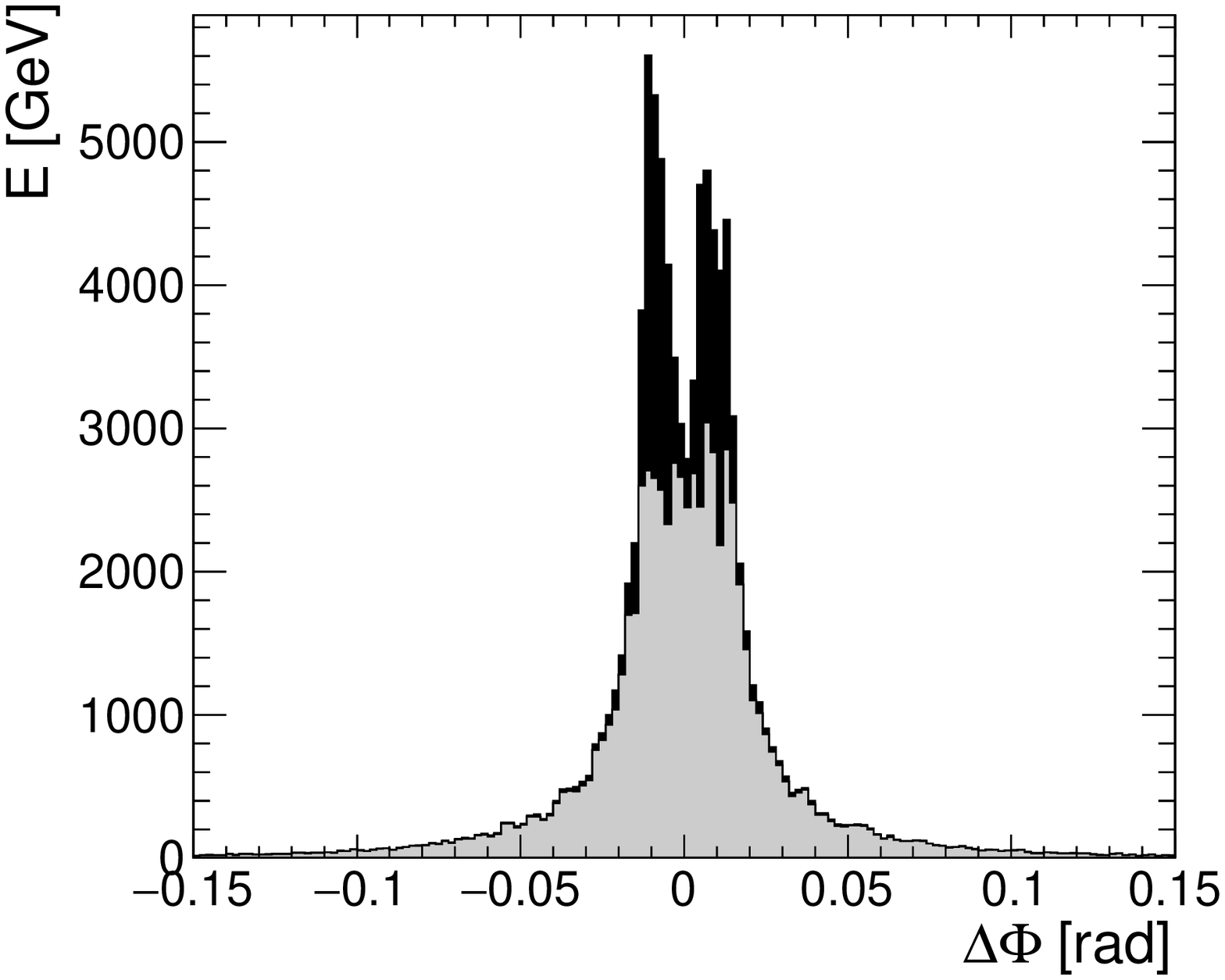}
  }
  \subfigure[$1 \times 1$~cm HCAL cells and $3 \times 3$~mm ECAL cells] {
   \includegraphics[width=0.45\textwidth]{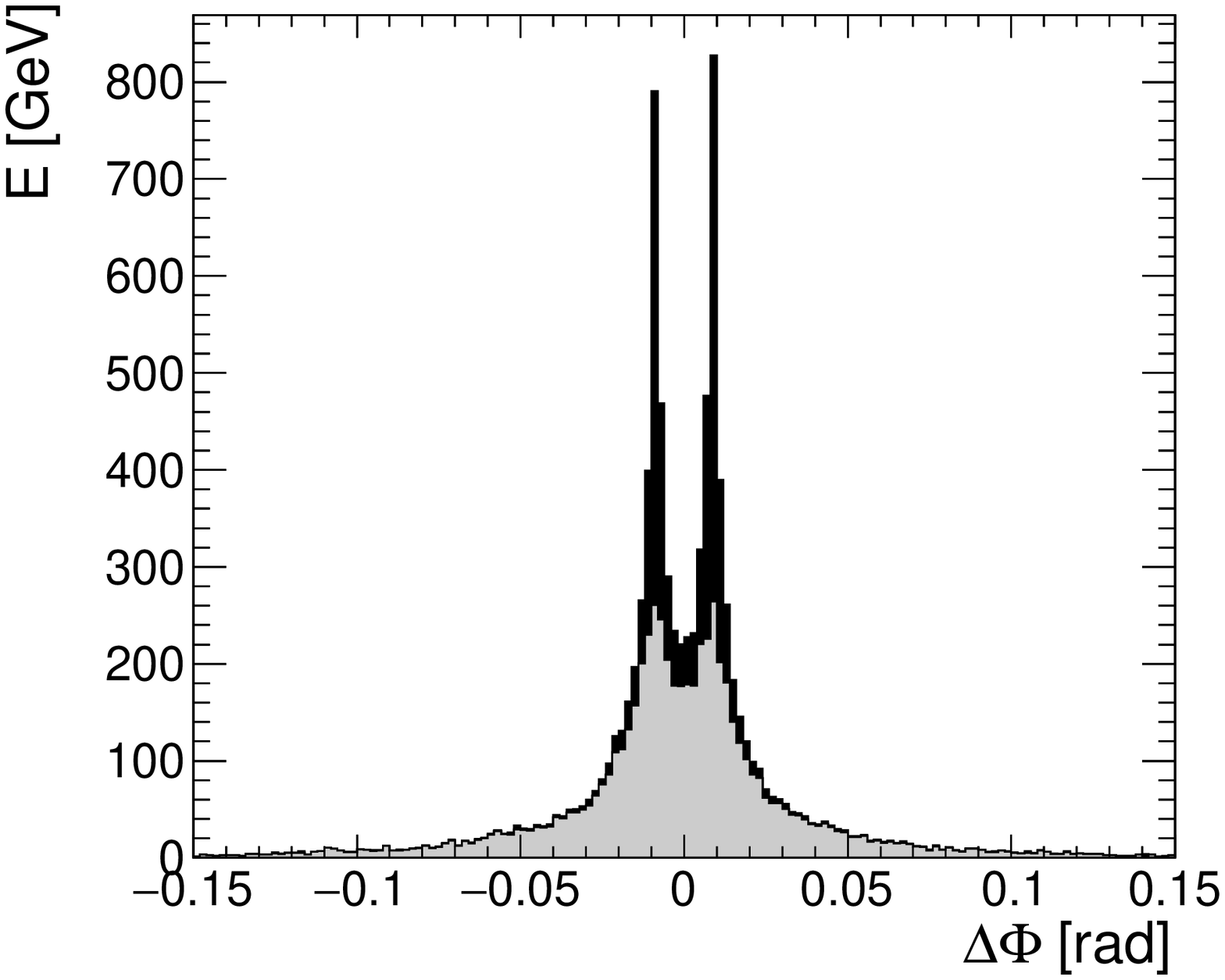}
   \includegraphics[width=0.45\textwidth]{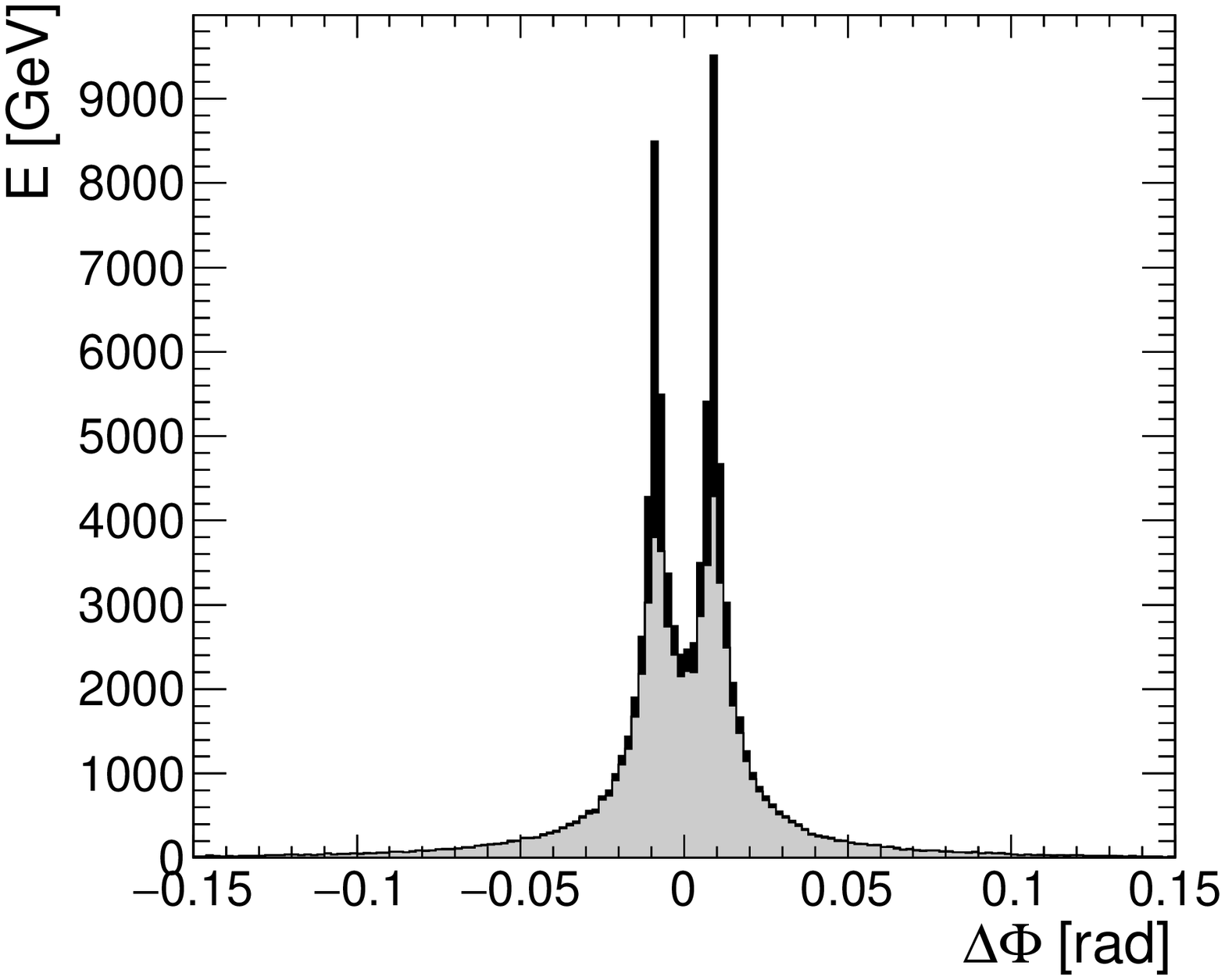}
 }
\caption{Azimuthal distribution of energy deposition for pair of incident $K^0_L$ particles at 100~GeV  (left) and 1000~GeV (right), with 
the angular separation of $\Delta \phi^{K}=0.018$~rad.
Electromagnetic calorimeter cells are indicated in black while hadronic calorimeter  cells  are indicated in gray.}
\label{fig:doublek2}
\end{figure}

\begin{figure} 
\centering
  \subfigure[$20 \times 20$~cm HCAL cells and $2 \times 2$~cm ECAL cells] {
  \includegraphics[width=0.45\textwidth]{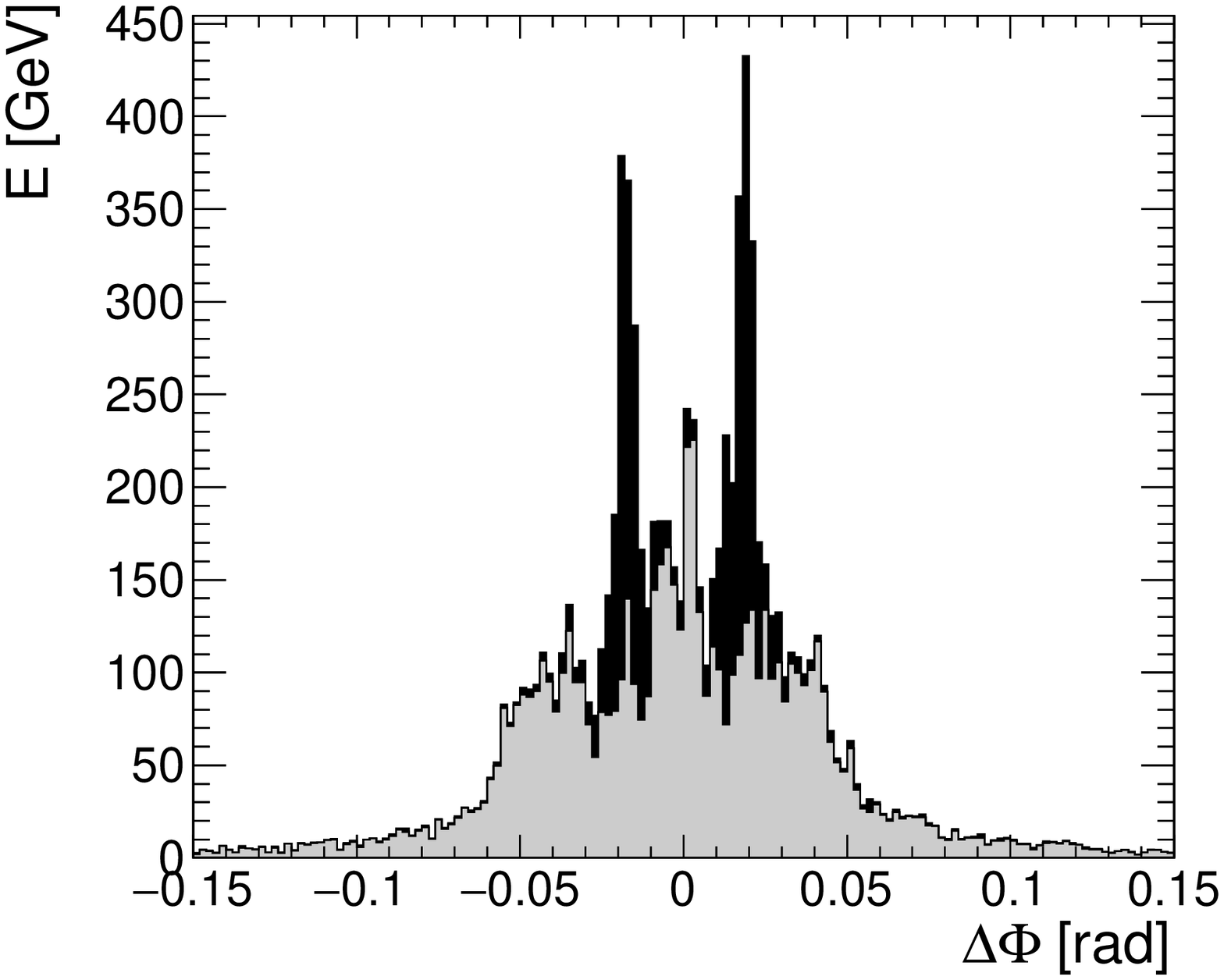}
  \includegraphics[width=0.45\textwidth]{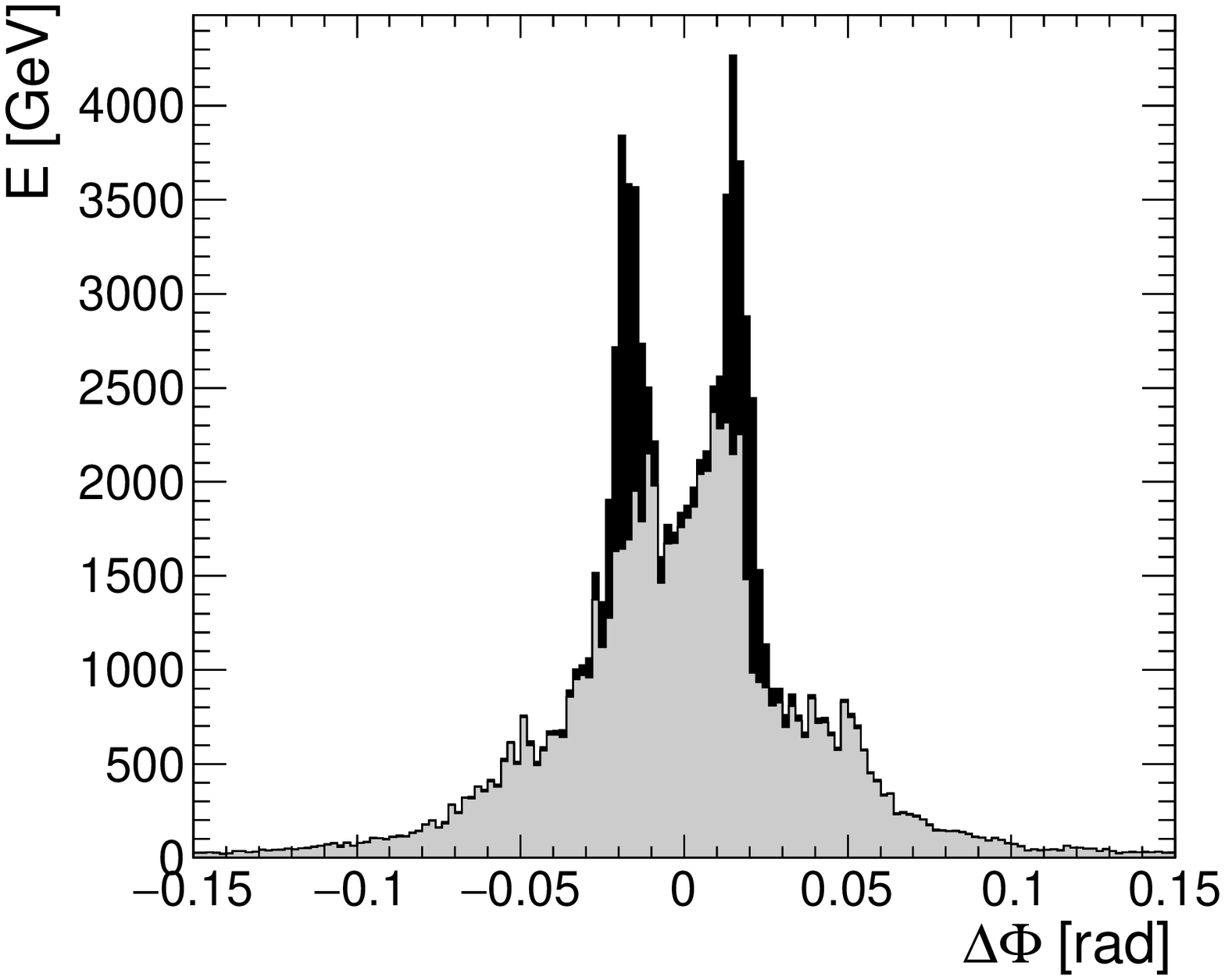}
  }
  \subfigure[$5 \times 5$~cm HCAL cells and $2 \times 2$~cm ECAL cells] {
  \includegraphics[width=0.45\textwidth]{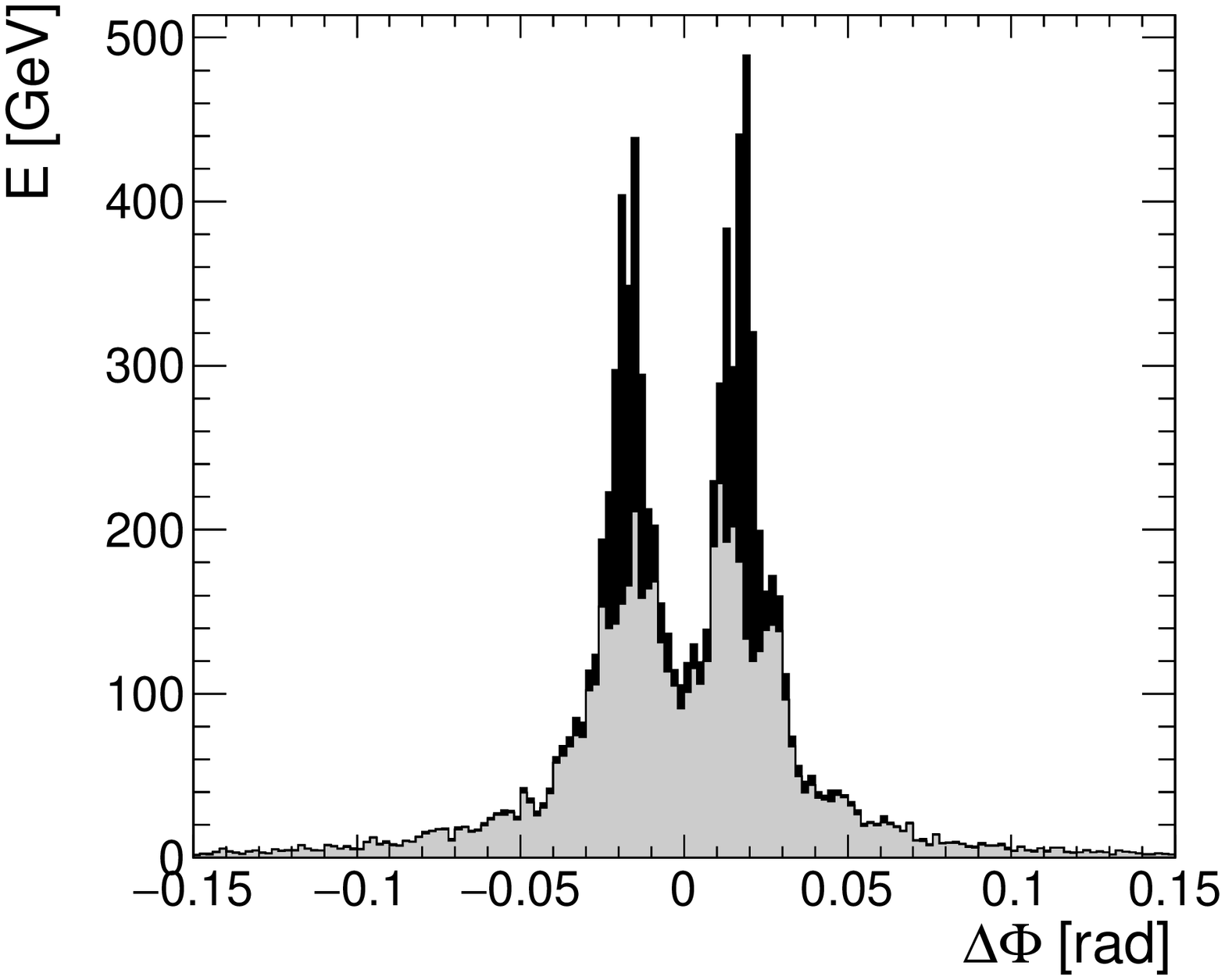}
  \includegraphics[width=0.45\textwidth]{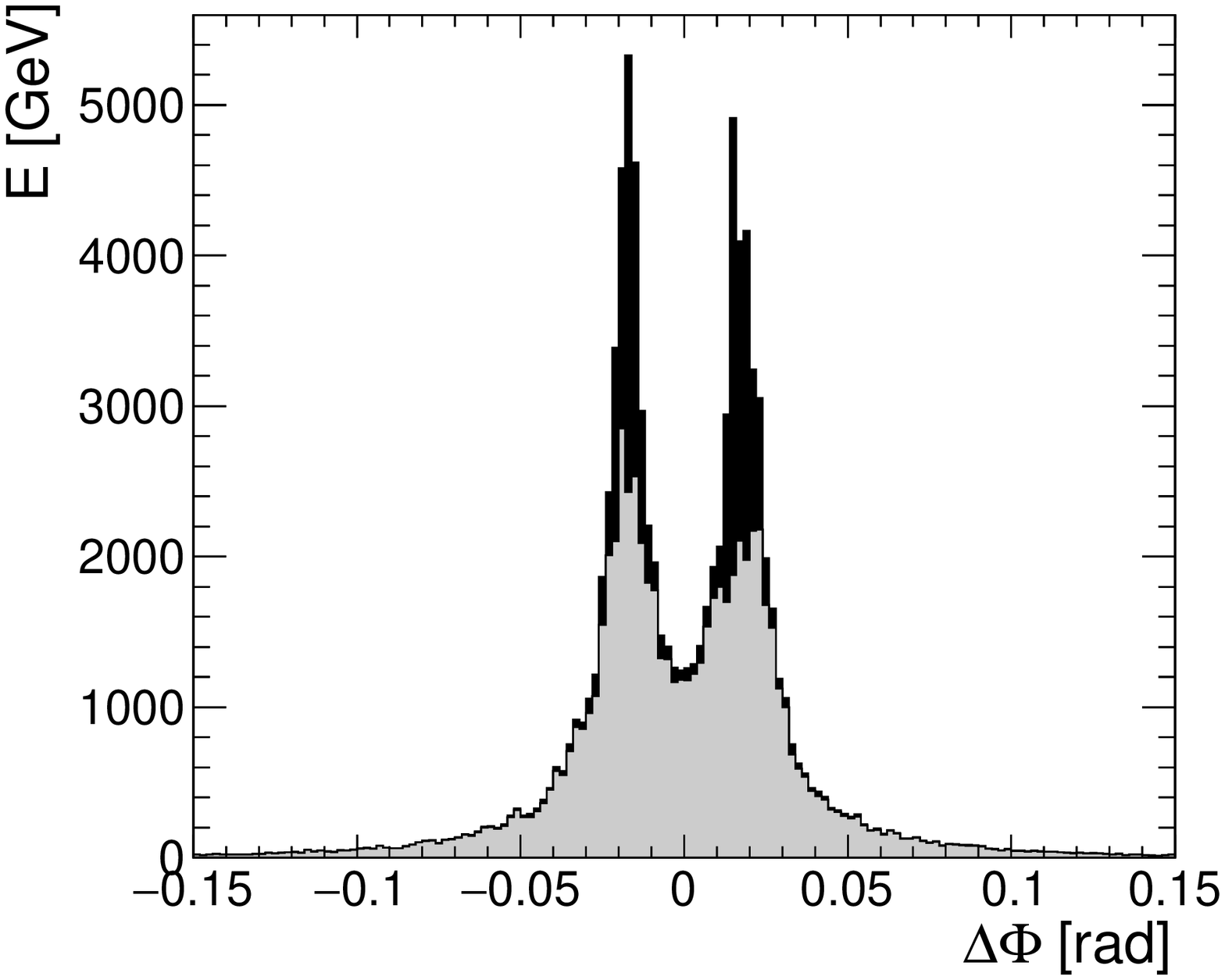}
  }
\caption{Azimuthal distribution of energy deposition for pair of incident $K^0_L$ particles at 100~GeV  (left) and 1000~GeV (right), with  the 
angular separation of $\Delta \phi^{K}=0.035$~rad.
Electromagnetic calorimeter cells are indicated in black while hadronic calorimeter  cells  are indicated in gray.}
\label{fig:doublek4}
\end{figure}

\begin{figure} 
\centering
  \subfigure[$K^0_L$  angular separation of 0.069~rad.] {
  \includegraphics[width=0.45\textwidth]{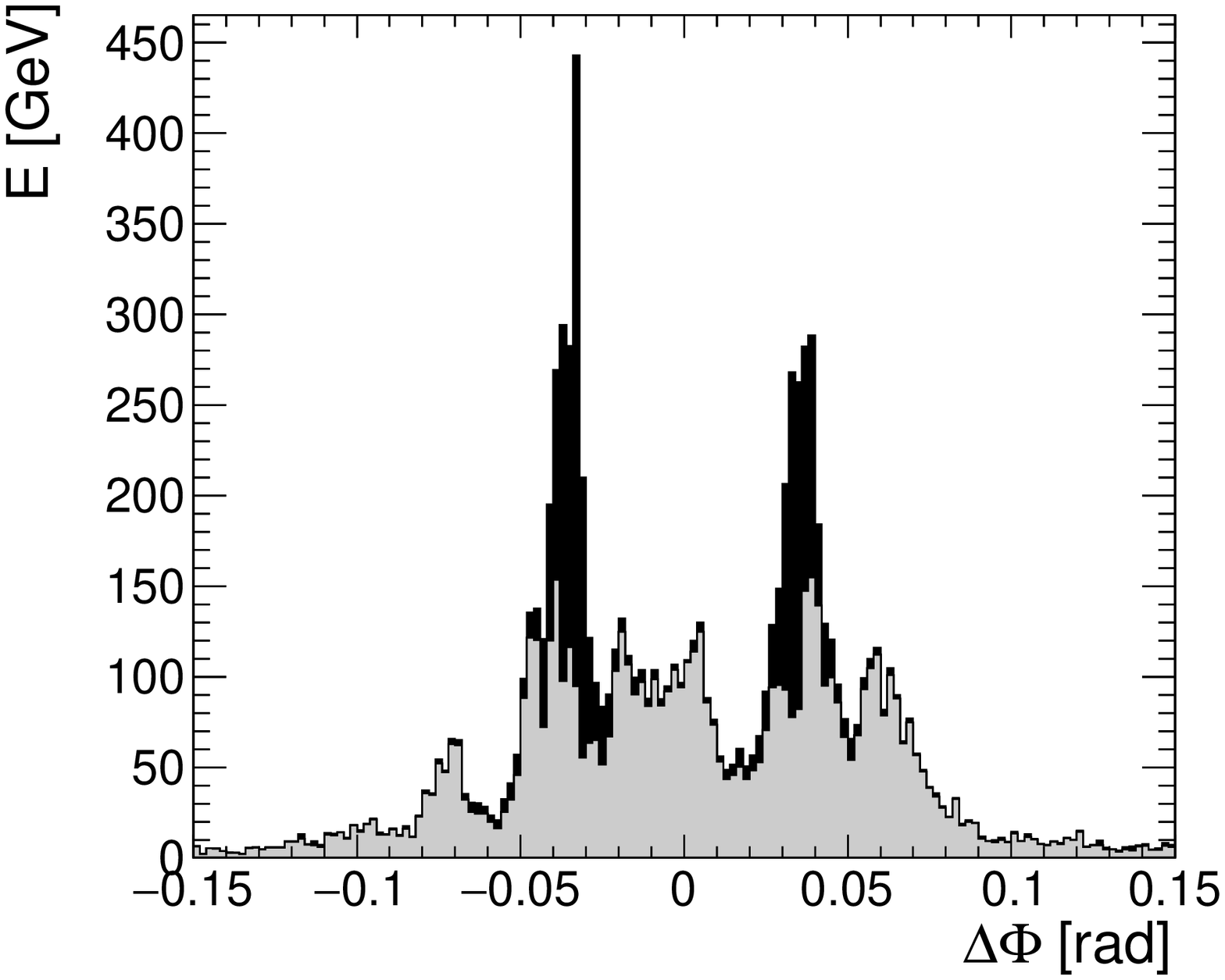}
  \includegraphics[width=0.45\textwidth]{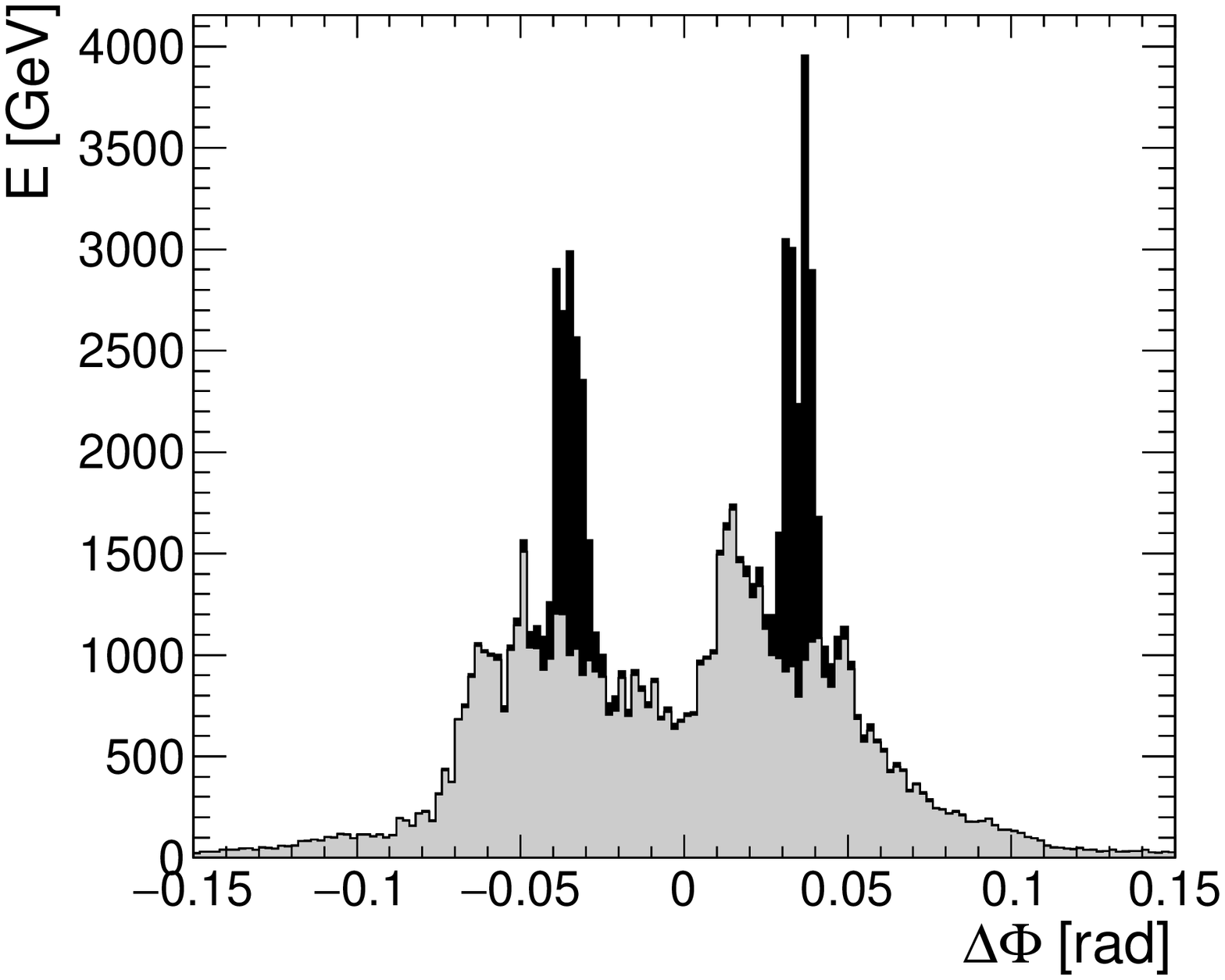}
  }
  \subfigure[$K^0_L$  angular separation of 0.104~rad.] {
  \includegraphics[width=0.45\textwidth]{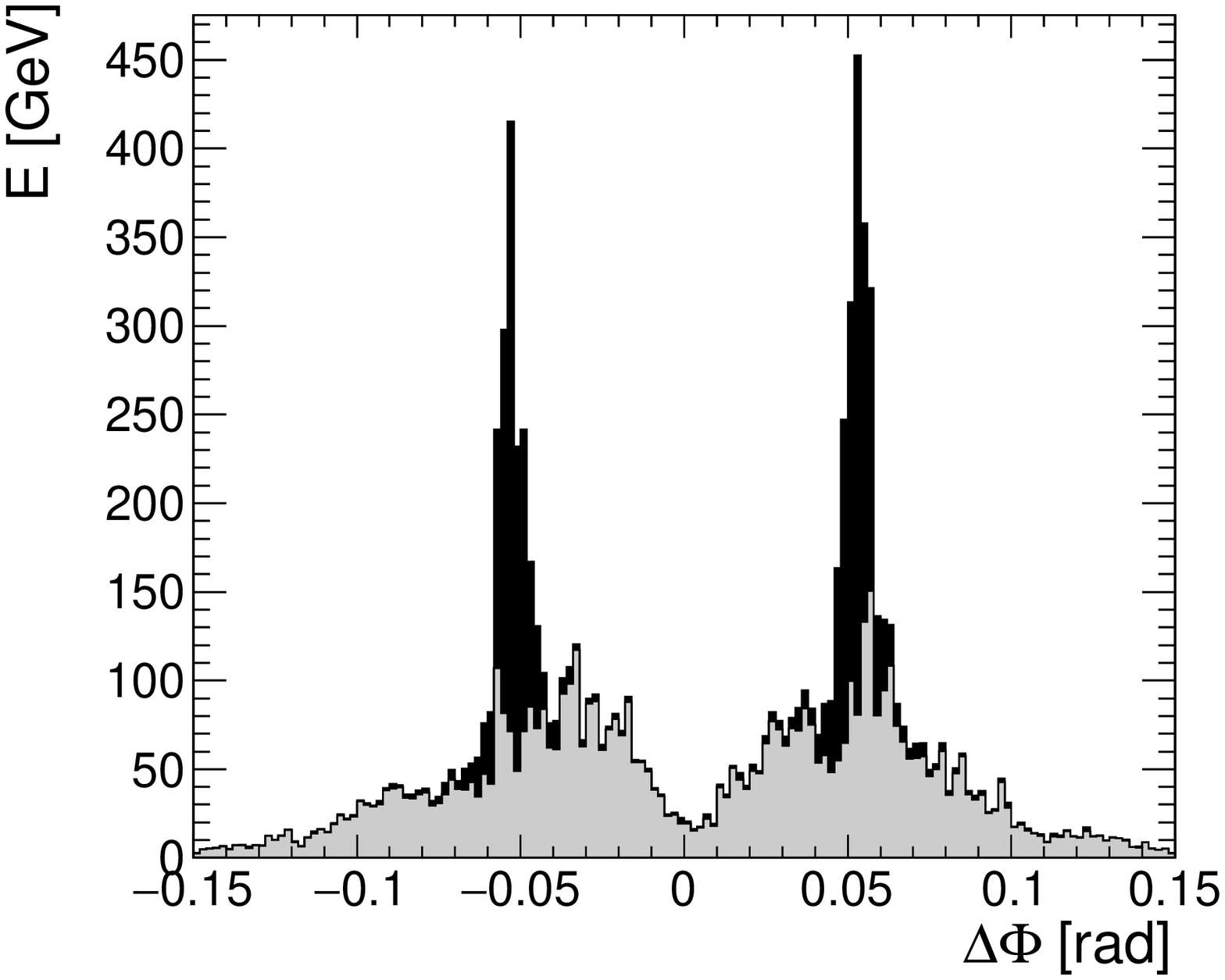}
  \includegraphics[width=0.45\textwidth]{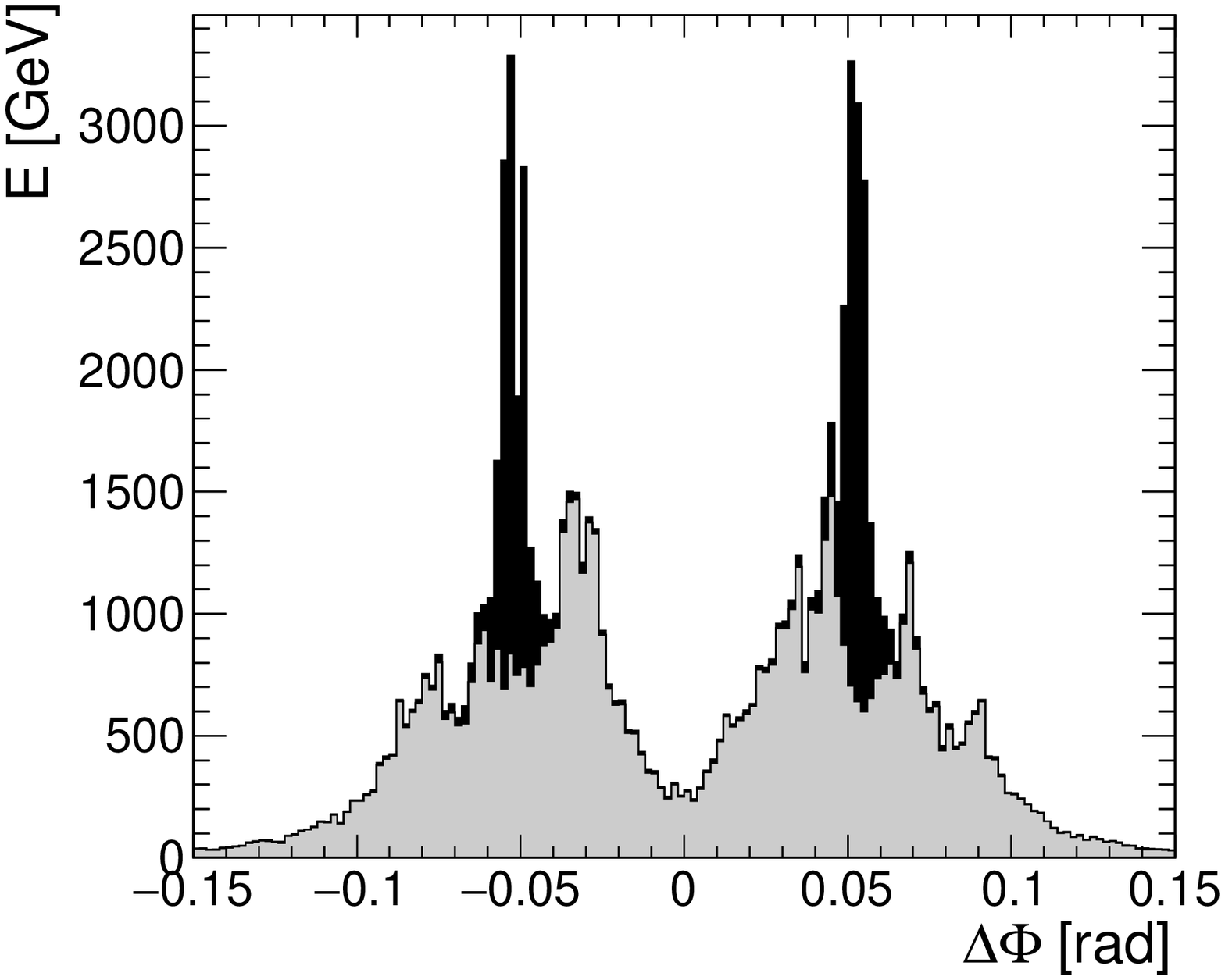}
  }
\caption{Azimuthal distribution of energy deposition for pair of incident $K^0_L$ particles at 100~GeV  (left) and 1000~GeV (right), for a 
 calorimeter with $20 \times 20$~cm HCAL cells and $2 \times 2$~cm ECAL cells. The distributions are shown for the 
incident $K^0_L$ angular separations of (a) 0.069~rad and (b) 0.104~rad. 
Electromagnetic calorimeter cells are indicated in black while hadronic calorimeter  cells are indicated in gray.}
\label{fig:doublek8}
\end{figure}

The ECAL and HCAL granularities were  changed to understand its impact on resolving the two $K^0_L$ showers.  
We plot the energy-weighted distribution of calorimeter 
cells in the azimuthal angle $\Phi$ for a fixed $K^0_L$ angular separation, as  shown in Fig.~\ref{fig:doublek1}-\ref{fig:doublek8} for $K^0_L$ angular separations 
$\Delta \phi^{K}$ in the range 
0.009 -- 0.104~rad and  for three calorimeter cell sizes.
The energies of cells were reconstructed form the sum of the corresponding calorimeter hits, after applying the sampling fractions. 
The ECAL cell size is  $ 2 \times 2$~cm when HCAL cell sizes are $20\times 20$~cm (configuration I) and $5\times 5$~cm (configuration II). 
 The ECAL cell size is $ 3 \times 3$~mm when the HCAL cell size is $1\times 1$~cm (configuration III).
The distributions are  integrated over 50 events.

 We find that as we improve granularity, the spatial resolving power of hadronic showers increases. To set the physical scale of the angular separation from boosted jets, 
 we consider a boosted $W$, $Z$ or Higgs boson with $p_T \sim 10$~TeV, producing decay quarks with a typical opening angle of 0.001~rad. Figure~\ref{fig:doublek1} shows the calorimeter 
 response of two $K^0_L$ particles with an angular separation of  $\Delta \phi^{K}=0.009$~rad. 
The hadrons are resolved in the ECAL of configuration III. Doubling the $K^0_L$ separation to $\Delta \phi^{K}=0.018$~rad  
  results in the hadrons being resolved in the ECAL of all three configurations, and the HCAL of configuration III, as shown in Fig.~\ref{fig:doublek2}. 
Further increasing  the 
$K^0_L$ separation to 0.035~rad results in them being resolved in the HCAL of configuration II, 
but not in configuration I (which is the current practice in HEP detectors such as CMS), 
as shown in Fig.~\ref{fig:doublek4}.
Finally, it is shown in Fig.~\ref{fig:doublek8} that the HCAL of configuration I can resolve hadrons with separation in the 0.07-0.1~rad range.

The energy-dependence of the spatial resolving power is also illustrated in these figures. 
In the hadron energy range of 100 -- 1000~GeV, the energy dependence is weak, but noticeable in the HCAL. 

These results can lead to an understanding of how well hadrons can be resolved on an event-by-event basis. 
 Figures~\ref{fig:doublek1}-\ref{fig:doublek8} indicate that one can resolve individual hadron  
showers at smaller separation than the nuclear interaction length (18~cm for steel, corresponding to angular separation of 0.08~rad  at a radius of 230~cm). 
The studies at the high energies presented here go beyond those  performed by the CALICE collaboration \cite{Adloff:2011ha}.

\newpage
\section{Summary}

A detector has been designed to study the physics performance at the energy scale of the FCC-hh or SppC collider.
The  concept of the SiFCC detector has been presented for the first time, as well as detailed characteristics relevant for multi-TeV physics. The 
performance of this detector has been illustrated using {\GEANTfour}-based Monte Carlo simulations of single incident particles of different species. 
It is shown that the tracking and calorimeter performance achieved 
with this initial general-purpose detector concept meets our expectations for the given technology choices. 
The resolution and reconstruction efficiency of single particles are well within the expected specifications. 
No significant leakage out the back of  the hadronic calorimeter is 
observed for single hadrons up to  33~TeV in transverse momentum.

Transverse momentum resolution of jets reconstructed from energy deposits in the calorimeters
is below  $1-2\%$ for jets with transverse momenta above $26$~TeV.
The jet response saturates at a value close to unity for $p_T^{\rm{jet}}>10$~TeV.
To date, this is the first estimate of jet resolution and response for jets in the $p_T^{\rm{jet}}$ range of tens of TeV using
a \GEANTfour-based simulation followed by realistic event reconstruction.

The study of double hadrons with small angular separation illustrates that hadronic showers of close-by particles can be resolved 
by using a high-granularity calorimeter. These  results 
 go beyond the studies performed for the ILC by the CALICE Collaboration in the context of particle flow algorithms. We belive that  our observations
help pave the way for using  high-granularity calorimeters for the reconstruction of multi-TeV jets and particles at future colliders.

Using this detector concept, various Monte Carlo event samples  
covering a wide range of physics processes have been made available in the HepSim data repository~\cite{Chekanov:2014fga}. 
In the future, optimized designs of this detector will be introduced after analysis of  simulated events of  
  the most important physics channels~\cite{Mangano:2016jyj,Contino:2016spe} for 100~TeV $pp$ collisions.

\section*{Acknowledgements}
We thank D.~Blyth, A.~Dotti, A.~Henriques, J.~Repond, J.~Proudfoot and A.~Ribon  for helpful discussions. 
This research was performed using resources provided by the Open Science Grid,
which is supported by the National Science Foundation and the U.S. Department of Energy's Office of Science. 
We gratefully acknowledge the computing resources provided on Blues, 
a high-performance computing cluster operated by the Laboratory Computing Resource Center at Argonne National Laboratory.
Argonne National Laboratory's work was supported by the U.S. Department of Energy, Office of Science under contract DE-AC02-06CH11357.
The Fermi National Accelerator Laboratory (Fermilab) is operated by Fermi Research Alliance, LLC under Contract No. DE-AC02-07CH11359 with the United States Department of Energy.

\newpage
\section*{References}

\bibliographystyle{elsarticle-num}
\def\bibname{\Large\bf References}
\def\refname{\Large\bf References}
\pagestyle{plain}
\bibliography{biblio}

\end{document}